\documentclass[english,twocolumn,floats,showpacs,amsmath,amssymb,prb]{revtex4}
\usepackage[T1]{fontenc}
\usepackage[latin9]{inputenc}
\usepackage{textcomp}
\usepackage{amsmath}
\usepackage{amssymb}
\usepackage{graphicx}

\makeatletter

\newcommand{\lyxmathsym}[1]{\ifmmode\begingroup\def\b@ld{bold}
  \text{\ifx\math@version\b@ld\bfseries\fi#1}\endgroup\else#1\fi}

\providecommand{\tabularnewline}{\\}

\@ifundefined{textcolor}{}
{%
 \definecolor{BLACK}{gray}{0}
 \definecolor{WHITE}{gray}{1}
 \definecolor{RED}{rgb}{1,0,0}
 \definecolor{GREEN}{rgb}{0,1,0}
 \definecolor{BLUE}{rgb}{0,0,1}
 \definecolor{CYAN}{cmyk}{1,0,0,0}
 \definecolor{MAGENTA}{cmyk}{0,1,0,0}
 \definecolor{YELLOW}{cmyk}{0,0,1,0}
 }

\usepackage{epsfig}

\usepackage{babel}

\makeatother

\usepackage{babel}
\begin{document}

\title{Quasi-harmonic approximation of thermodynamic properties of ice Ih,
II, and III}

\author{R. Ramírez$^{1,a)}$, N. Neuerburg\textsuperscript{1}, M.-V. Fernández-Serra\textsuperscript{2},
and C. P. Herrero\textsuperscript{1} \let\oldthefootnote\thefootnote\global\long\def\thefootnote{{a)}}
 \footnotetext{Electronic mail: ramirez@icmm.csic.es}\let\thefootnote\oldthefootnote }

\affiliation{\textsuperscript{1}Instituto de Ciencia de Materiales de Madrid
(ICMM), Consejo Superior de Investigaciones Científicas (CSIC), Campus
de Cantoblanco, 28049 Madrid, Spain}

\affiliation{\textsuperscript{2}Department of Physics and Astronomy, Stony
Brook University, Stony Brook, New York 11794-3800, USA}

\date{{\today}}
\begin{abstract}
Several thermodynamic properties of ice Ih, II, and III are studied
by a quasi-harmonic approximation and compared to results of quantum
path integral and classical simulations. This approximation allows
to obtain thermodynamic information at a fraction of the computational
cost of standard simulation methods, and at the same time permits
studying quantum effects related to zero point vibrations of the atoms.
Specifically we have studied the crystal volume, bulk modulus, kinetic
energy, enthalpy and heat capacity of the three ice phases as a function
of temperature and pressure. The flexible q-TIP4P/F model of water
was employed for this study, although the results concerning the capability
of the quasi-harmonic approximation are expected to be valid independently
of the employed water model. The quasi-harmonic approximation reproduces
with reasonable accuracy the results of quantum and classical simulations
showing an improved agreement at low temperatures ($T$< 100 K). This
agreement does not deteriorate as a function of pressure as long as
it is not too close to the limit of mechanical stability of the ice
phases. 
\end{abstract}


\maketitle

\section{Introduction\label{sec:intro}}

Computer simulations of water and ice remain a very active research
field after the first pioneering works more than 40 years 
ago.\citep{barker69,rahman71}
Despite the impressive progress achieved in the development of total
energy computational models, simulation methods, and computer architectures,
we are still far from having a standard model to simulate water at
the atomistic level. In fact, water molecules are currently simulated
with various degrees of sophistication. In the order of increasing
complexity, one finds water models as point particles with short-range
anisotropic interactions,\citep{moore11} as rigid molecules,\citep{jorgensen05,vega11}
as flexible molecules with either harmonic\citep{paesani06} or anharmonic
OH bonds,\citep{habershon09} or as flexible polarizable molecules.\citep{stern01,fanourgakis08}
Accurate characterization of the electronic structure of the molecule
and its chemical reactivity is only possible when \textit{ab initio}
methods, like density functional theory (DFT),\citep{benoit02,chen03,marivi04,marivi06,morrone08,yoo09}
are used. Within this field, the development and application of new
functionals specially designed to treat van der Waals interactions
is a focus of recent interest.\citep{dion04,murray11,wang11,santra11}

An additional aspect in the simulation of water phases is performing
the calculation of thermodynamic properties either in a classical
or a quantum limit. Evidence of the relevance of quantum effects related
to nuclear motions in water phases is provided by measurements of
isotope effects, that would vanish in the classical limit. As an example
the melting point of ice Ih under normal conditions is shifted by
3.8 K in deuterated ice, and by 4.5 K in tritiated ice. Interestingly,
the density of ice Ih shows an inverse isotope effect, i.e., D$_{2}$O
ice is expanded with respect to H$_{2}$O ice,\citep{rottger12,rottger94}
whose microscopic origin has been recently revealed.\citep{pamuk12}

The path integral (PI) formulation of statistical mechanics is the
most applied method to treat quantum nuclear motion in 
water.\citep{kuharski85,mahoney01,stern01,chen03,hernandezpena04,shiga05,paesani07,morrone08,habershon09,mcbride09,vega10,ramirez10,herrero11,herrero11b,ramirez11}
However, the computational cost of this method is much larger than
that of its classical counterpart, and for this reason many equilibrium
properties of water and ice have not been yet studied by quantum simulations.
A prominent example is the complex phase diagram of ice, with fifteen
different phases. To date, the simulation of relevant parts of this
phase diagram has been only accessible to classical simulations using
effective rigid water models.\citep{sanz04,vega09} The computational
overhead of the quantum PI simulations is caused by the need to work
with multiple replicas of the system. Several approximations are available
that help to reduce it. For the case of empirical point charge models
of water, the ring polymer contraction scheme allows for a significant
increase of computational efficiency.\cite{markland08} In the case
of \textit{ab initio} methods the number of replicas can be reduced
dramatically by means of an appropriate generalized Langevin 
equation.\cite{ceriotti11}

The quasi-harmonic approximation (QHA) has been occasionally applied
to study thermodynamic properties of ice phases. Here the collective
vibrations are described by a set of harmonic oscillators and thus
the partition function can be expressed as an analytic function of
the crystal volume and the temperature. An advantage of this approximation
is that all equilibrium thermodynamic properties can be derived 
straightforwardly.
This implies that the computational effort of the QHA is only a very
small fraction of that required for a PI simulation of an ice phase,
and that this fraction is independent of the chosen model potential.
Further advantages of the QHA are the absence of statistical errors,
as opposed to any classical or quantum simulation, and the possibility
to account for finite size effects by a Brillouin zone integration
of the phonon dispersion curves, rather than by increasing the size
of the cell.

The QHA in combination with \textit{ab initio} DFT has allowed the
explanation of the inverse isotope effect in the crystal volume of
ice Ih at atmospheric pressure.\citep{pamuk12} Also the negative
thermal expansion of ice Ih at low temperatures has been studied by
the QHA,\citep{tanaka01} as well as the elastic moduli and mechanical
stability of the hydrogen ordered ice VIII.\citep{tse99} In addition,
the mechanical stability of ice Ih under pressure has been studied
by this approximation.\citep{tse99b} One conclusion of these results
is that the QHA seems to be reasonably realistic for ice phases. However,
a systematic study of its potential as a predictive tool is still
missing.

The purpose of this work is to check the ability of the QHA to predict
thermodynamic properties of ice phases. We have analyzed three different
phases (ices Ih, II, and III) in order to have a broad field of comparison.
Our aim here is to compare PI results to the expectations of a much
simpler QHA. We have used the point charge, flexible q-TIP4P/F model
to investigate several thermodynamic properties as a function of both
pressure and temperature.\citep{habershon09} This model has been
recently employed in PI investigations of isotope effects,\citep{ramirez10,herrero11}
proton kinetic energies,\citep{ramirez11} and pressure effects in
water and ice Ih.\citep{herrero11b} We expect that the validity of
the QHA will be to a large extent independent of the employed potential
model.

The layout of the manuscript is as following. Computational details
of the numerical simulations and QHA of the ice phases Ih, II, and
III are given in Sec. \ref{sec:Computational-procedures}. The phonon
density of states (DOS) and the Gr\"uneisen constants calculated with
the q-TIP4P/F model are presented in Sec. \ref{sec:Q-tip4p/f-ice-phonons}.
Comparison of QHA and simulation results for various pressures ($P$)
and temperatures ($T$) are presented in Sec. \ref{sec:Results}.
Specifically, we study the volume ($V$), bulk modulus ($B$), kinetic
energy ($K$), enthalpy ($H$), and heat capacity ($C_{p})$. Also
the comparison to experimental data is presented whenever available.
The paper closes with the conclusions.

\section{Computational procedures\label{sec:Computational-procedures}}

In this Section we summarize some relevant computational details concerning
the ice cells, the PI simulations and the QHA.

\subsection{Simulation cells}

Ice Ih displays proton disorder, i.e., while the oxygen atoms occupy
fixed lattice positions, the water molecules display random orientations
with the constraint that the resulting H-bond network must be compatible
with the Bernal-Fowler ice rules.\citep{pauling35} These rules imply
that each oxygen is tetrahedrally coordinated to two H atoms by strong
OH covalent bonds ($\sim$490 kJ/mol) and to two additional H atoms
by weaker H-bonds ($\sim$25 kJ/mol). Ice II is an ordered phase where
the molecular orientation is fully determined by the crystal symmetry.
Ice III displays partial proton disorder, i.e., the occupancy of 2/3
of the crystallographic H sites deviates from 50\% and it is found
between 35 and 65\%.\citep{lobban00} Proton disordered structures
with nearly zero dipole moment were generated for ice Ih and III by
a simple MC algorithm designed to explore molecular orientations that
obey the ice rules.\citep{buch98} In line with recent computer simulations
a full proton disordered structure was assumed for ice III,\citep{haberschon11,vega09,cogoni11}
as this simplification seems to have only a minor effect in the phase
diagram of ice.\citep{vega09} The largest ice cells employed in our
study contain $N=$ 288 molecules for ice Ih, and 324 molecules for
ices II and III. The ice Ih cell was orthorhombic with parameters
$(4\mathbf{a}_{1},3\sqrt{3}\mathbf{a}_{1},3\mathbf{a}_{3})$, with
$(\mathbf{a}_{1},\mathbf{a}_{3})$ being the standard hexagonal lattice
vectors of ice Ih,\citep{hayward87} while ice II and ice III were
studied by $3\times3\times3$ supercells of the crystallographic cell,
which belong to the rhombohedral and the tetragonal crystal systems,
respectively.\citep{kamb71,lobban00} Simulations were conducted also
for smaller simulation cells to check the convergence of the results
with the system size. In the following the computational conditions
employed for the PIMD simulations are presented.

\subsection{Path integral simulations}

In the PI formulation of statistical mechanics the quantum partition
function is calculated through a discretization of the integral representing
the density matrix. This discretization leads to a suggestive picture:
with respect to its equilibrium thermodynamic properties, the quantum
system results to be \textit{isomorphic} to a classical one. However,
this isomorphism does not apply to the dynamic (time dependent) properties
of the system. Specifically, the classical isomorph is constructed
by a simple substitution of each quantum particle (here, atomic H
and O nucleus in the water molecule) by a ring polymer of $L$ classical
particles (beads), connected by harmonic springs with a temperature-
and mass-dependent force constant. Details of this practical simulation
method can be found elsewhere.\citep{feynman72,gillan88,ceperley95,chakravarty97}
Equilibrium properties in the classical isomorph can be derived by
a classical molecular dynamics (MD) algorithm, that has the advantage
against a Monte Carlo method of being more easily parallelizable,
an important fact for efficient use of modern computer architectures.
Effective reversible integrator algorithms to perform PIMD simulations
have been described in Refs. \onlinecite{ma99,tu02,tu98,tuckerman93}.
Ref. \onlinecite{ma96} introduces useful methods to treat full
cell fluctuations and multiple time step integration. All simulations
were done using originally developed software and parallelization
was implemented by the MPI library.\citep{pacheco97} The potential
energy of each replica of the system is calculated in parallel by
a different processor. Other interesting optimization schemes, not
used in this work, are based on ring polymer contraction techniques.\cite{markland08}

\begin{figure}
\vspace{-1.2cm}
\includegraphics[width= 9cm]{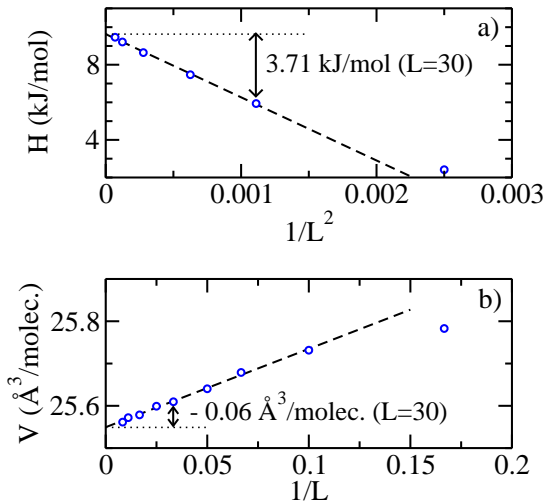}
\vspace{-0.8cm}
\caption{ Results for the convergence of the enthalpy (a) and
volume (b) of ice II with the number of beads ($L$) employed in the
PIMD simulations. The constant shifts that correct the finite $L$
results derived with the condition $LT=$ 6000 K are explicitly shown.
The simulations were performed at $T=$200 K and $P=$1 atm.}
\label{fig:L_convergence}
\end{figure}

PIMD simulations in the isothermal-isobaric $NPT$ ensemble ($N$
being the number of water molecules) were conducted for ice Ih, II,
and III at temperatures up to 300 K and pressures between -1 and 2
GPa. We have employed the point-charge, flexible q-TIP4P/F model,
that was originally parameterized to provide the correct liquid structure,
diffusion coefficient, and infrared absorption frequencies in quantum
simulations.\citep{habershon09} Periodic boundary conditions were
applied to the simulation cell and the Ewald method was employed to
calculate the long range electrostatic potential and atomic forces.
Expected averages were derived in runs of $10^{6}$ MD steps (MDS)
using a time step of $\Delta t=0.2$ fs. The system equilibration
was conducted in runs of $10^{5}$ MDS. To have a nearly constant
precision as a function of temperature, the number of beads $L$ was
set as the integer number closest to fulfill the relation $LT=6000$
K, i.e., at 200 K the number of beads was $L=30$. The classical limit
is easily achieved by setting $L$=1 in the PIMD algorithm. The largest
error associated to the finite number of beads is caused by the highest
energy vibrations, i.e., the intramolecular OH stretching modes. These
modes remain essentially in their ground state as their energy quantum
($\hbar\omega_{OH}$) is several times larger ($\sim16$) than the
thermal energy ($k_{B}T$) at the highest studied temperature. Thus
we expect this error to be a nearly $T$ (and $P$) independent shift.
The shift has been estimated for the quantities of interest (volume,
enthalpy, and kinetic energy) by a series of $NPT$ simulations of
the three ice phases using different numbers of beads at $T$= 200
K and $P$= 1 atm. The $L\rightarrow\infty$ limit was extrapolated
from the best linear fit in either the variable $1/L^{2}$ or $1/L$.
In Fig. \ref{fig:L_convergence} the result of this extrapolation
is represented for both the enthalpy and volume of ice II. The constant
shifts that correct the thermal averages derived with the relation
$LT=6000$ K are 3.71 kJ/mol for the enthalpy and 1.89 kJ/mol for
the kinetic energy. The corresponding correction shift for the volume
amounts to -0.03 $\textrm{\AA}^{3}$/molecule for ice Ih and III,
and -0.06 $\textrm{\AA}^{3}$/molecule for ice II. Additional computational
conditions in the simulations were identical to those reported in
Ref. \onlinecite {ramirez10}.

\subsection{Quasi-harmonic approximation}

\subsubsection{Basic equations}

The quasiharmonic approximation employed here is based on the following
three assumptions:

$a)$ the simulation cell, defined by the vectors ($\mathbf{a}_{1},\mathbf{a}_{2},\mathbf{a}_{3})$,
is considered to scale isotropically with the crystal volume, $V$,
as 
\begin{equation}
\mathbf{a}_{i}(V)=(V/V_{ref})^{1/3}\mathbf{a}_{i,ref}\;,i=1,2,3\;,.\label{eq:a_i}
\end{equation}
 $V_{ref}$ is the volume of the reference cell 
($\mathbf{a}_{1,ref},\mathbf{a}_{2,ref},\mathbf{a}_{3,ref})$
that minimizes the potential energy at a certain chosen pressure,
$P_{ref}$, where the ice phase is mechanically stable.

$b)$ the lattice vibrations are described by harmonic oscillators
of wavenumber $\omega_{k}$, with $k$ combining the phonon branch
index and the wave vector within the Brillouin zone.

$c)$ the wavenumbers $\omega_{k}$ depend on the volume of the crystal,
which may change with temperature and pressure. However, for a given
volume the wavenumbers $\omega_{k}$ are constants (independent of
$T$ and $P$) as in a harmonic approximation.

With these assumptions, the Helmholtz free energy of the ice cell
with $N$ molecules in a volume $V$ and at temperature $T$ is given
by
\begin{equation}
F(V,T)=U_{S}(V)+F_{v}(V,T)-TS_{H}\;,\label{eq:f_v_t}
\end{equation}
 where $U_{S}(V)$ is the static zero-temperature classical energy,
i.e., the minimum of the potential energy when the volume of the ice
cell is $V$. The entropy $S_{H}$ is related to the disorder of hydrogen,
it vanishes for ice ordered phases as ice II. $S_{H}$ was estimated
by Pauling for fully disorder phases using simple counting schemes
of allowed water orientations \citep{pauling35}
\begin{equation}
S_{H}=Nk_{B}\ln\frac{3}{2}\;.
\end{equation}
 $F_{v}(V,T)$ is the vibrational contribution to $F$,
\begin{equation}
F_{v}(V,T)=\sum_{k}\left(\frac{\hbar\omega_{k}}{2}+\frac{1}{\beta}\ln\left[1-\exp\left(-\beta\hbar\omega_{k}\right)\right]\right)\;.\label{eq:fv_q}
\end{equation}
 Here $\beta$ is the inverse temperature: $1/k_{B}T.$ This expression
for $F_{v}$ is valid for quantum harmonic oscillators. If one is
interested in the classical limit of the QHA the vibrational contribution
changes to
\begin{equation}
F_{v,cla}(V,T)=\sum_{k}\frac{1}{\beta}\ln\left(\beta\hbar\omega_{k}\right)\;.
\end{equation}
 The terms $U_{S}(V)$ and $S_{H}$ remain the same for either a quantum
or classical limit. The calculation of those thermodynamic properties
derived from the partition function (as the Gibbs free energy, enthalpy,
internal energy, kinetic energy, state equation, thermal expansion,
bulk modulus, heat capacity, etc.) can be easily performed with the
help of Eq. (\ref{eq:f_v_t}). For example, the Gibbs free energy,
$G(T,P),$ can be derived by seeking for the volume, $V_{min}$, that
minimizes the function $F(V,T)+PV$, as
\begin{equation}
G(P,T)=F(V_{min},T)+PV_{min}\;.
\end{equation}

\subsubsection{Direct implementation\label{sub:Direct-implementation}}

The direct implementation of the QHA implies the following steps:

$a)$ determination of the ice reference cell by an energy minimization
where both water molecules and cell parameters are simultaneously
optimized. Both the shape of the reference cell and its volume $V_{ref}$
are obtained in this step.

$b)$ a set of cell volumes $\left\{ V_{i}\right\} $ of interest
is selected.

$c)$ given a cell volume $V_{i}$, the simulation cell is derived
by the scaling of the reference cell as shown by Eq. (\ref{eq:a_i}).
The water molecules are then relaxed to their minimum potential energy
and the crystal phonons are calculated and tabulated. This tabulation
allows us the calculation of $F(V_{i},T)$ for the set of volumes
$\left\{ V_{i}\right\} $ by means of Eq (\ref{eq:f_v_t}).

$d)$ for the purpose of finding the minimum of $F(V_{i},T)$ as a
function of $V$ (or alternatively, the minimum of $G=F+PV$), it
is convenient to perform a polynomial fit of $F(V_{i},T)$ and then
look for the minimum of the fitted function.

This implementation requires, for each of the studied volumes $V_{i}$,
a crystal structure minimization and the calculation of the corresponding
crystal phonons. Typically we have selected a set of 50 different
$V_{i}$ for each ice phase and chosen a 5th degree polynomial as
fitting function to find the minimum of $F(V,T)$ as a function of
$V$. Using an empirical potential this implementation is straightforward
and does require a little amount of computer time even for simulation
cells containing several hundreds of water molecules.

The crystal phonon calculation has been performed by the small-displacement,
or frozen phonon method.\citep{kresse95,alfe01} The basic principle
is to compute the force constants between atom pairs numerically deriving
the (analytic) forces. Atoms are displaced a small, but finite, amount
from their perfect-lattice positions, and all the atomic forces generated
by this displacement are calculated. Given that the displacement is
small, the force constant describes the proportionality between displacements
and forces. The atomic displacement employed in this work is $\delta x=10^{-6}$
$\textrm{\AA}$ along each cartesian direction. Although we have used
a flexible water model, the method can be also applied to rigid water
models by allowing molecular displacements associated with small translations
and rotations of the rigid units. See Ref. {[}\onlinecite{venkataraman70}{]}
for a full account of the calculation of the external phonon modes
associated to rigid molecules.

The QHA results have been derived by a $\Gamma$ ($\mathbf{k}=0$)
sampling of the Brillouin zone of the ice simulation cell. Although
this is an adequate approximation, given the large size of the employed
cells, its main justification is that we are interested in the comparison
to PIMD simulations. Note that the application of periodic boundary
conditions in the simulation implies that all atomic images move in
phase, and this corresponds exactly to the spatial symmetry of the
$\mathbf{k}=0$ phonons.

\subsubsection{A simpler implementation}

For calculations based on non-empirical potentials of water, e.g.,
in \emph{ab initio} DFT theory, it may be computationally convenient
to reduce the number of energy minimizations and phonon calculations
by using the phonon-dependent Gr\"uneisen parameters defined as\citep{ashcroft76}
\begin{equation}
\gamma_{k}=-\left(\frac{\partial\ln\omega_{k}}{\partial\ln V}\;\right)_{V_{ref}}.\label{eq:g}
\end{equation}
 This equation can be integrated assuming that $\gamma_{k}$ is a
volume independent constant to obtain the volume dependence of $\omega_{k}$
as
\begin{equation}
\omega_{k}(V)\approx\omega_{k}(V_{ref})\times\left(\frac{V}{V_{ref}}\right)^{-\gamma_{k}}\;.\label{eq:g_1}
\end{equation}
 A linear expansion of the previous relation, i.e., taking the derivative
$(\partial\omega_{k}/\partial V)$ as a constant independent of $V$,
leads to the alternative expression
\begin{equation}
\omega_{k}(V)\approx\omega_{k}(V_{ref})\times\left(1-\frac{V-V_{ref}}{V_{ref}}\gamma_{k}\right)\;.\label{eq:g_2}
\end{equation}
 The last two relations allow us to calculate the function $F_{v}(V,T)$
in Eq. (\ref{eq:fv_q}) from a tabulation of the values $(\omega_{k},\gamma_{k})$
obtained for the reference cell. We have compared the direct implementation
of the QHA (see Sect. \ref{sub:Direct-implementation}) with the simpler
implementation of Eqs. (\ref{eq:g_1}) and (\ref{eq:g_2}) by calculating
the function $V(T)$ at $P=0$. We find that Eq. (\ref{eq:g_2}) works
slightly better than Eq. (\ref{eq:g_1}) for ice Ih, while for ice
II and III the opposite behavior was observed. A recent QHA study
of the thermal expansion of ice Ih using DFT was based on Eq. (\ref{eq:g_2}).\citep{pamuk12}
Note that the numerical calculation of the Gr\"uneisen constants requires
at least two energy minimizations and phonon calculations (at $V_{ref}$
and at an additional volume in the proximity of $V_{ref}$). Therefore
the computational cost of this simpler implementation is reduced by
about a factor of 25 in comparison to the previous direct implementation
of the QHA. In the next Section the direct implementation of the QHA
is applied to the study of ice phases.

\begin{table}
\caption{Volume, potential energy, and simulation cell parameters of the
reference
cells of ice Ih, II, and III obtained by energy minimizations with
the q-TIP4P/F model at $P_{ref}=0.$ The direct implementation of
the QHA implies a phonon calculation at 50 different volumes uniformly
distributed in the interval {[}$V_{min},$$V_{max}${]}.}
\label{tab:ref}
\vspace{4mm}
\begin{tabular}{lr@{\extracolsep{0pt}.}lr@{\extracolsep{0pt}.}lr@{\extracolsep{0pt}.}l}
\hline
 & \multicolumn{2}{c}{Ih} & \multicolumn{2}{c}{II} &
\multicolumn{2}{c}{III}\tabularnewline
\hline
$N$ (molecules/cell)  & \multicolumn{2}{c}{288} &
\multicolumn{2}{c}{324} & \multicolumn{2}{c}{324}\tabularnewline
$V_{ref}$ ($\lyxmathsym{\AA}\text{\textthreesuperior/molec.)}$  & 30 &
96  & 24 & 14  & 24 & 99\tabularnewline
$U_{S,ref}$ (kJ/mol)  & $\quad$-61 & 98$\quad$  & $\quad$-60 & 84$\quad$
& $\quad$-60 & 86$\quad$\tabularnewline
$\triangle U_{S,ref}$ (kJ/mol)  & -1 & 15  & \multicolumn{2}{c}{0} & -0
& 02\tabularnewline
$a_{1,ref}$ ($\textrm{\AA})$  & 17 & 78  & 22 & 98  & 19 &
68\tabularnewline
$a_{2,ref}$ ($\textrm{\AA})$  & 23 & 09  & 22 & 98  & 19 &
77\tabularnewline
$a_{3,ref}$ ($\textrm{\AA})$  & 21 & 76  & 22 & 98  & 20 &
80\tabularnewline
$\alpha$ ($\text{\textsuperscript{0}}$)  & 90 & 0  & 113 & 2  & 89 &
9\tabularnewline
$\beta$ ($\text{\textsuperscript{0}}$)  & 90 & 0  & 113 & 2  & 89 &
9\tabularnewline
$\gamma$ ($\text{\textsuperscript{0}}$)  & 90 & 0  & 113 & 2  & 89 &
8\tabularnewline
$V_{min}$($\lyxmathsym{\AA}\text{\textthreesuperior/molec.)}$  & 29 & 47
& 21 & 75  & 22 & 48\tabularnewline
$V_{max}$ ($\lyxmathsym{\AA}\text{\textthreesuperior/molec.)}$  & 35 &
05  & 27 & 31  & 28 & 22\tabularnewline
\hline
\end{tabular}
\end{table}

\section{Q-tip4p/f ice phonons\label{sec:Q-tip4p/f-ice-phonons}}

Reference cells for the QHA have been derived at $P_{ref}$=0 for
the three ice phases by energy minimizations implying optimization
of both atomic positions and simulation cell. Optimized simulation
cells as well as the corresponding volumes ($V_{ref})$ and potential
energies ($U_{S,ref})$ are summarized in Tab. \ref{tab:ref}. The
direct implementation of the QHA involved the calculation of the vibrational
modes for 50 different volumes uniformly distributed in the interval
{[}$V_{min},$$V_{max}${]} (see Tab. \ref{tab:ref}). Note that the
largest volume per molecule corresponds to ice Ih (28\% and 24 \%
larger than those of ices II and III, respectively). By setting the
potential energy of ice II as the zero of the energy scale, the structure
of ice Ih results more stable by $\triangle U_{S,ref}$=-1.15 kJ/mol,
a value in good agreement to that of -1.17 kJ/mol reported in Ref.
\onlinecite{haberschon11}. However, for ice III we find $\triangle U_{S,ref}=$
-0.02 kJ/mol, significantly different from the previously reported
result of -0.1 kJ/mol.\citep{haberschon11} We have checked that this
difference is a consequence of the disorder of hydrogen in ice III.
Support for this is given by the equilibrium potential energies $U_{S,ref}$
of a series of five randomly generated ice III structures. These energies
scatter in an energy window of 0.2 kJ/mol, a value more than two times
larger than the energy difference between our results and the one
Ref. \onlinecite{haberschon11}. Interestingly this effect is much
smaller in disordered ice Ih, where the energy fluctuations computed
for five randomly generated structures are reduced by more than an
order of magnitude to about 0.01 kJ/mol.

\begin{figure}[!t]
\vspace{-0.3cm}
\includegraphics[width= 9cm]{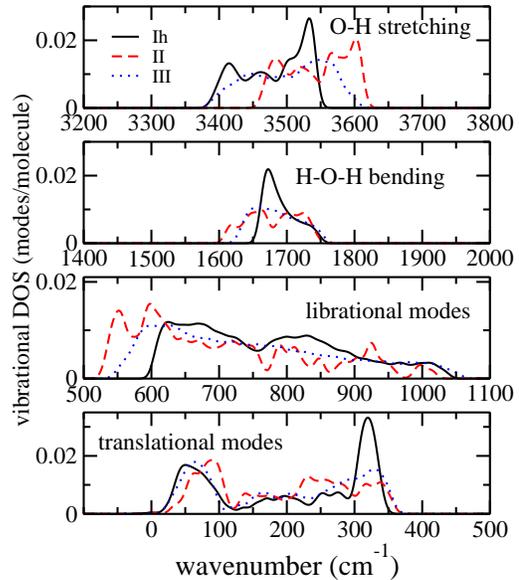}
\vspace{-0.8cm}
\caption{Vibrational density of states obtained for the
reference
cells of ice Ih, II, and III with the q-TIP4P/F water model. The
harmonic
modes correspond to a $\Gamma$ sampling of the Brillouin zone and
each mode was plotted as a normalized Gaussian with a width of 10
cm$^{-1}$. }
\label{fig:w}
\end{figure}

\begin{figure}
\vspace{-0.6cm}
\includegraphics[width= 9cm]{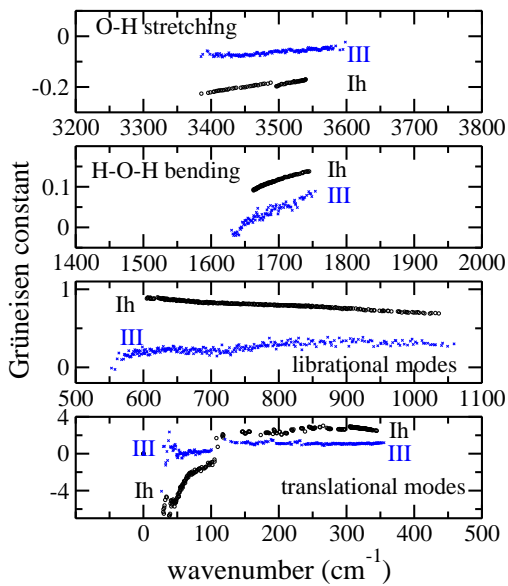}
\vspace{-0.8cm}
\caption{Gr\"uneisen constants of ice Ih and III calculated
for
the q-TIP4P/F model. Results for ice II lie in-between ice Ih and
III and are not shown for clarity. }
\label{fig:gru}
\end{figure}

The vibrational DOS calculated with the q-TIP4P/F model for the reference
cells of the three ice phases are shown in Fig.\ref{fig:w}. The wavenumbers
of the vibrational modes are separated into four groups comprising
3$N$ translational, 3$N$ librational, $N$ bending, and 2$N$ stretching
modes. Translational and librational modes are related to the intermolecular
interactions and can be differentiated not only by their respective
frequency windows but also by their effective masses. The translational
modes display a {}``heavy'' molecular mass while the librational
modes display a {}``light'' hydrogen mass, a fact that can be experimentally
observed by the frequency shifts found after isotopic substitution
of either O or H atoms. Within a harmonic approximation this shift
scales as the squared root of the effective mass. The intramolecular
modes (bending and stretchings) display also a {}``light'' hydrogen
mass. It is interesting to observe the anticorrelation between librational
and stretching modes in the ice phases: the lower the frequency of
librational modes, the higher the frequency of the stretchings modes.\citep{libowitzky99,bratos09}
The calculated Gr\"uneisen constant for ice Ih and III are displayed
in Fig. \ref{fig:gru}.

A fingerprint of the vibrational structure of the ice phases is presented
in Tab. \ref{tab:w}, that summarizes the result of averaging the
complete set of $9N$ wavenumbers calculated for the ice reference
cell into 9 groups of $N$ modes taken from a list ordered by increasing
wavenumber. The resulting average frequencies and Gr\"uneisen constants
reveal significant differences between the ice phases. The negative
Gr\"uneisen constant associated to the $N$ translational modes of lowest
energy is the origin of the negative thermal expansion experimentally
found in ice Ih at low temperature.\citep{rottger94} For ice II the
averaged Gr\"uneisen constant of these modes is also negative but much
smaller in absolute value, while for ice III is positive. Negative
Gr\"uneisen parameters are also found for the $2N$ stretching vibrations
with absolute values in the order Ih>II>III. We note that the Gr\"uneisen
constants of the stretching vibrations in the q-TIP4P/F model are
a factor of two smaller than those obtained in the \textit{ab initio
}DFT study of ice Ih.\citep{pamuk12} These larger DFT absolute values
are crucial to correctly describe the experimental inverse isotope
effect found in the crystal volume of normal and deuterated ice Ih,
\citep{pamuk12} an effect that is not reproduced by the q-TIP4P/F
model.\citep{herrero11}

\begin{table*}
\caption{Average of the 9$N$ harmonic modes obtained for the reference
cells
of the ice phases with the q-TIP4P/F model into groups of $N$ modes.
The average wavenumbers and Gr\"uneisen constants are a fingerprint
of the vibrational structure of each ice phase.}
\label{tab:w}
\vspace{4mm}
\begin{tabular}{r|ccc|r@{\extracolsep{0pt}.}lr@{\extracolsep{0pt}.}lr@{\extracolsep{0pt}.}l}
\hline
 &  & $\left\langle \omega_{k}\right\rangle $ (cm$^{-1}$)  &  &
\multicolumn{2}{c}{} & \multicolumn{2}{c}{$\left\langle
\gamma_{k}\right\rangle $} & \multicolumn{2}{c}{}\tabularnewline
modes  & Ih  & II  & III  & \multicolumn{2}{c}{Ih} &
\multicolumn{2}{c}{II} & \multicolumn{2}{c}{III}\tabularnewline
\hline
$N$ translational  & 60  & 76  & 63  & $\;$-3 & 32  & $\quad$-0 &
14$\quad$  & 0 & 14\tabularnewline
$N$ translational  & 223  & 194  & 209  & 2 & 47  & 2 & 44  & 1 &
14\tabularnewline
$N$ translational  & 323  & 302  & 319  & 2 & 74  & 2 & 67  & 1 &
10\tabularnewline
\hline
$N$ librational  & 650  & 574  & 613  & 0 & 86  & 0 & 78  & 0 &
20\tabularnewline
$N$ librational  & 758  & 673  & 726  & 0 & 81  & 0 & 76  & 0 &
24\tabularnewline
$N$ librational  & 904  & 859  & 907  & 0 & 75  & 0 & 64  & 0 &
32\tabularnewline
\hline
$N$ bending  & 1691  & 1678  & 1684  & 0 & 11  & 0 & 10  & 0 &
03\tabularnewline
$N$ stretching  & 3439  & 3506  & 3453  & -0 & 20  & -0 & 15  & -0 &
07\tabularnewline
$N$ stretching  & 3522  & 3587  & 3549  & -0 & 18  & -0 & 13  & -0 &
05\tabularnewline
\hline
\end{tabular}
\end{table*}

\section{QHA Results\label{sec:Results}}

In this Section we evaluate the performance of the QHA in the description
of ice Ih, II, and III by comparing it to PIMD simulation results.
Our study is focused on the temperature and pressure dependence of
several ice properties as crystal volume, bulk modulus, kinetic energy,
enthalpy, and heat capacity. Both quantum and classical limits have
been studied, and compared to experimental results whenever they are
available.

\subsection{Thermal expansion}

\begin{figure}
\vspace{-0.6cm}
\includegraphics[width= 9cm]{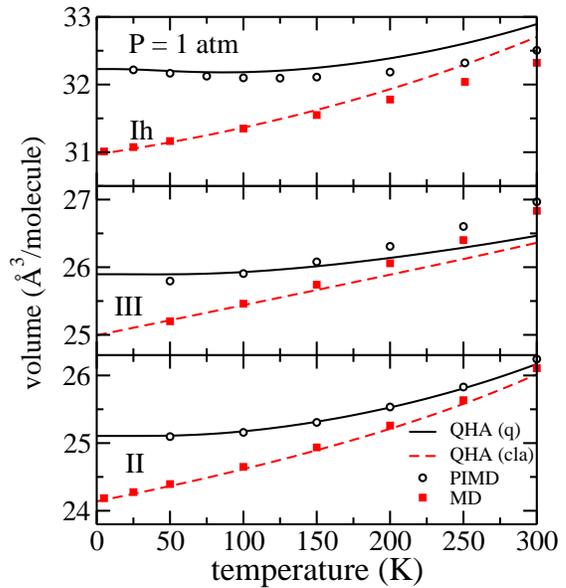}
\vspace{-0.8cm}
\caption{ Thermal expansion of ice Ih, II, and III at
atmospheric
pressure calculated with the q-TIP4P/F water model. The QHA results
(lines) are compared to PIMD and MD simulations (symbols). Both quantum
(q) and classical (cla) limits are shown. Error bars of the simulation
results are less than the symbol size.}
\label{fig_v_t_pi}
\end{figure}

The temperature dependence of the crystal volume at atmospheric pressure
derived from the QHA is compared to PIMD results in Fig. \ref{fig_v_t_pi}.
Both quantum and classical limits are shown for the three ice phases.
The QHA reproduces the simulation results rather accurately up to
temperatures of 100 K. At higher temperatures anharmonic effects not
included in the QHA cause deviations of different sign in the ice
Ih and III curves, while for ice II the agreement with the PIMD data
is good up to the highest studied temperature. The QHA estimation
of the zero point expansion (i.e. the volume increase with respect
to the classical limit at $T$=0) of the q-TIP4P/F model amounts to
4\% for ice Ih and II, and to 3.6\% for ice III.

\begin{figure}
\vspace{0.5cm}
\includegraphics[width= 9cm]{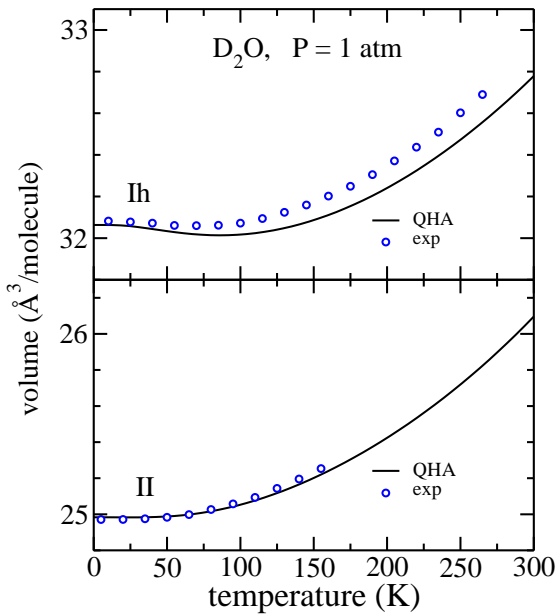}
\vspace{-0.8cm}
\caption{Temperature dependence of the volume of D$_{2}$O
ice
Ih and II at atmospheric pressure. The QHA results (lines) are compared
to the experimental data (circles) of ice Ih (Ref.
\onlinecite{rottger94})
and II (Ref. \onlinecite{fortes05}).}
\label{fig:v_t_exp}
\end{figure}

The comparison of the QHA result to available experimental data\citep{rottger94,fortes05}
at atmospheric pressure is presented in Fig. \ref{fig:v_t_exp} for
deuterated ices Ih and II. The overall agreement is good, in particular
for ice II. At low temperatures there appears a negative thermal expansion
in both experimental and q-TIP4P/F results of ice Ih. However, for
ice II this effect is not appreciable at the scale of the figure.
This behavior of ice Ih and ice II can be rationalized by the differences
in the average Gr\"uneisen constant of the $N$ translational modes
of lowest frequency (see Tab. \ref{tab:w} for values calculated for
the H$_{2}$O ices). The negative thermal expansion of ice Ih is a
stringent test of the water model. An interesting comparison of several
effective potentials reveals that the negative thermal expansivity
of ice Ih is succesfully reproduced by the TIP4P model. However this
effect is absent in the TIP5P potential and only slightly visible
with the ST2 model.\cite{koyama04}

\subsection{State equation at 100 K}

\begin{figure}
\vspace{-0.6cm}
\includegraphics[width= 9cm]{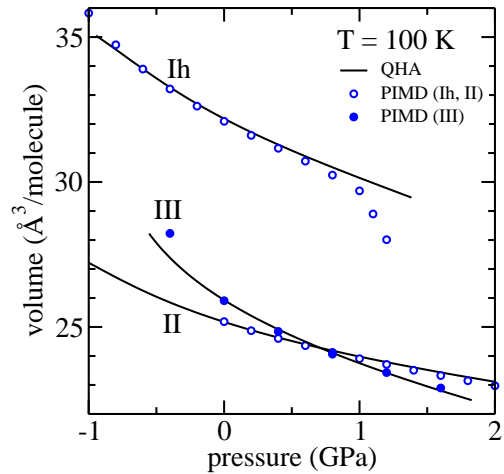}
\vspace{-0.8cm}
\caption{ Equation of state of ice Ih, II, and III at 100
K.
QHA results (lines) are compared to PIMD simulations (symbols). All
results were derived with the q-TIP4P/F water model. Error bars of
the simulation results are less than the symbol size.}
\label{fig:v_p_pi}
\end{figure}

The pressure dependence of the volume of the ice phases is presented
in Fig. \ref{fig:v_p_pi} at temperature of 100 K. The QHA shows good
agreement with PIMD results. The sudden drop of the PIMD data of ice
Ih at high pressures is due to the proximity of the spinodal pressure
at about 1.2 GPa where ice Ih becomes mechanically unstable and collapses
into a high-density amorphous (HDA) ice in simulations with the q-TIP4P/F
model.\citep{herrero11b} Interestingly this limit of mechanical stability
of ice Ih is also captured by the QHA. By increasing the hydrostatic
pressure the QHA predicts a compressed ice Ih volume characterized
by the appearance of soft phonons that reach a vanishing vibrational
frequency at a pressure slightly below 1.4 GPa.

The best overall agreement between PIMD and QHA results is found for
H ordered ice II. In the case of ice III a significant deviation is
found at negative pressures of about -0.5 GPa, but for positive pressures
the QHA provides accurate results.

\begin{figure}
\vspace{-1.7cm}
\includegraphics[width= 9cm]{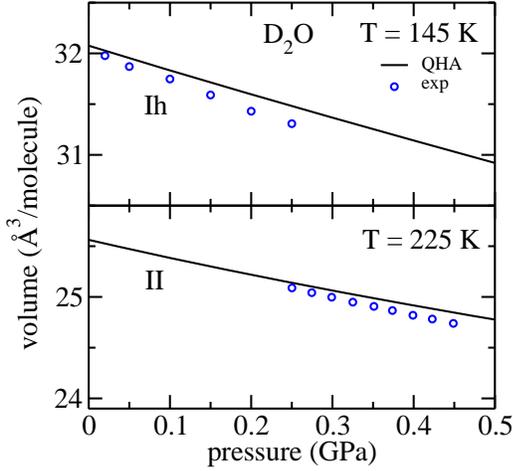}
\vspace{-0.8cm}
\caption{ Equation of state of D$_{2}$O ice at 145 K (ice
Ih)
and 225 K (ice II). The QHA (lines) is compared to experimental data
(circles) of ice Ih (Ref. \onlinecite{strassle05}) and II (Ref.
\onlinecite{fortes05}).}
\label{fig:v_p_exp}
\end{figure}

The experimental results for D$_{2}$O ice at 145 K (ice Ih)\citep{strassle05}
and 225 K (ice II)\citep{fortes05} are compared to the QHA expectation
in Fig. \ref{fig:v_p_exp}. The experimental data show the pressure
window where each phase is stable at the studied temperature. Note
the large volume difference between the two ice phases. The QHA predicts
reasonable results for both phases specially in the low pressure region
of each phase. The main difference to the experimental data is seen
by the slope of the $V(P)$ curve, i.e., the QHA using the q-TIP4P/F
model overestimates the hardness of both ices Ih and II (it predicts
a larger bulk modulus).

\subsection{Bulk modulus}

\begin{figure}
\vspace{-0.8cm}
\includegraphics[width= 9cm]{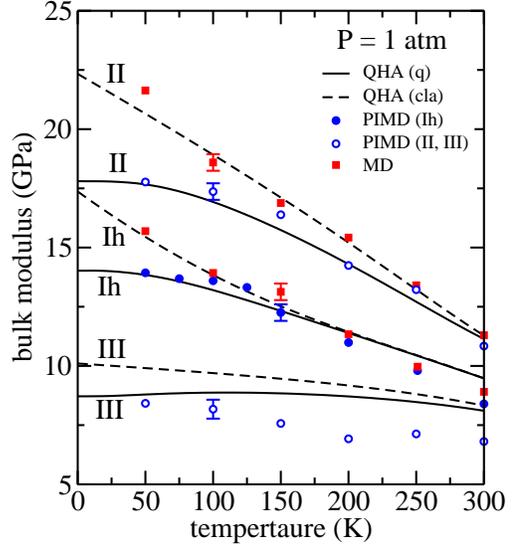}
\vspace{-0.8cm}
\caption{ Temperature dependence of the bulk modulus of
ice Ih,
II, and III derived with the q-TIP4P/F mode at atmospheric pressure.
The QHA results in the quantum and classical limits (lines) are compared
to PIMD and classical MD simulations, respectively (symbols). Typical
error bars in the simulation results are shown at 100 K (ices II and
III) and 150 K (ice Ih).}
\label{fig:b_t_pi}
\end{figure}

The temperature dependence of the bulk modulus of the three studied
ice phases is presented in Fig. \ref{fig:b_t_pi}. The bulk modulus
in the MD simulations is calculated from the fluctuation formula of
the cell volume in the $NPT$ ensemble.\citep{herrero11b} At any
given temperature the order of increasing bulk modulus (less compressibility)
is: ice III < Ih < II. The QHA of ice II provides results in excellent
agreement with quantum PIMD and classical MD simulations. Also for
ice Ih we find that the QHA provides good results. A worse agreement
is found for ice III where the QHA predicts a bulk modulus somewhat
larger than the PIMD results. Classical and quantum results above
100 K are not very different from each other specially for ice III
and Ih. The statistical error of the bulk modulus in the MD simulations
is large. The classical MD results for ice III have not been represented
as its statistical error was of the order of its difference to the
PIMD data.

\begin{figure}
\vspace{-1.5cm}
\includegraphics[width= 9cm]{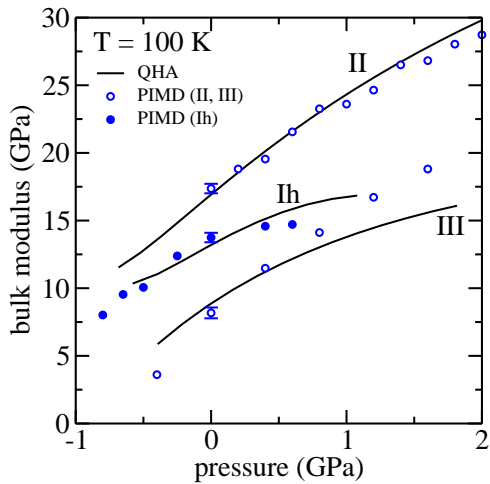}
\vspace{-0.8cm}
\caption{ Pressure dependence of the bulk modulus of ice
Ih,
II, and III calculated with the q-TIP4P/F model at 100 K. QHA results
(lines) are compared to PIMD simulations (symbols). The estimated
error bars of the simulation results are shown at $P=$ 0.}
\label{fig:b_p_pi}
\end{figure}

The pressure dependence of the bulk modulus of ice Ih, II, and III
calculated with the QHA is compared in Fig. \ref{fig:b_p_pi} with
the results of PIMD simulations at 100 K. The increase with pressure
of the bulk modulus predicted by the QHA agrees rather accurately
with the PIMD data in the case of ice II and ice Ih. For ice III we
find again a worse agreement between QHA and MD simulation results.
Nevertheless, the largest deviations between both sets of data are
found either at negative pressures or at positive ones larger than
0.35 GPa. This value is the coexistence pressure experimentally found
at 100 K between ice III and the higher pressure phase ice V.\citep{dunaeva10}

\subsection{Kinetic energy\label{sub:Kinetic-energy}}

\begin{figure}
\vspace{-0.1cm}
\includegraphics[width= 9cm]{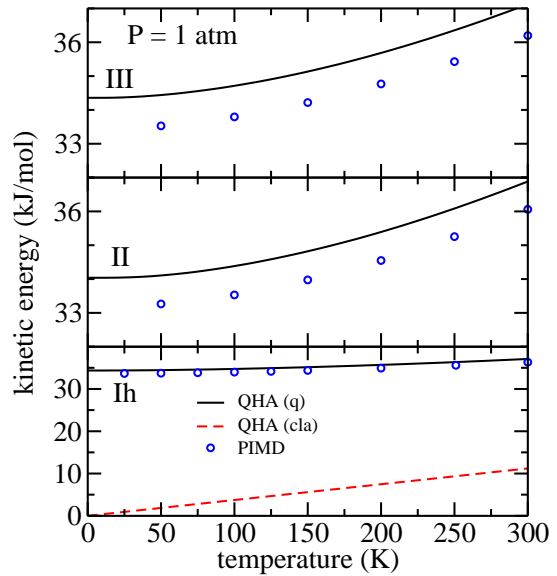}
\vspace{-0.8cm}
\caption{ Kinetic energy of ice Ih, II, and III at
atmospheric
pressure calculated with the q-TIP4P/F water model. The QHA results
(full lines) are compared to PIMD (symbols). For ice Ih both quantum
(q) and classical (cla) limits are shown. Error bars of the simulation
results are smaller than the symbol size.}
\label{fig:k_t}
\end{figure}

The internal energy is given by the sum of potential and kinetic energy
contributions. Within the QHA both energy terms must be identical
as a consequence of the virial theorem. A comparison of the QHA and
PIMD estimations of the kinetic energy, $K$, at atmospheric pressure
is presented in Fig. \ref{fig:k_t} as a function of temperature.
For ice Ih we compare also the quantum results to the classical limit
(broken line), that amounts to $k_{B}T/2$ per degree of freedom (equipartition
principle). The QHA estimation of the zero point kinetic energy of
ice Ih is 34.4 kJ/mol. Nearly the same value is obtained for ice III,
while for ice II we get a reduced zero-point energy energy of 34.0
kJ/mol. This difference implies that ice II is stabilized with respect
to ice Ih and ice III due to its lower zero point energy. The comparison
of QHA and PIMD data shows that the QHA overestimates the value of
$K$ by about 0.8 kJ/mol in the three ice phases and that this shift
is nearly temperature independent. 

\begin{figure}
\vspace{-1.8cm}
\includegraphics[width= 9cm]{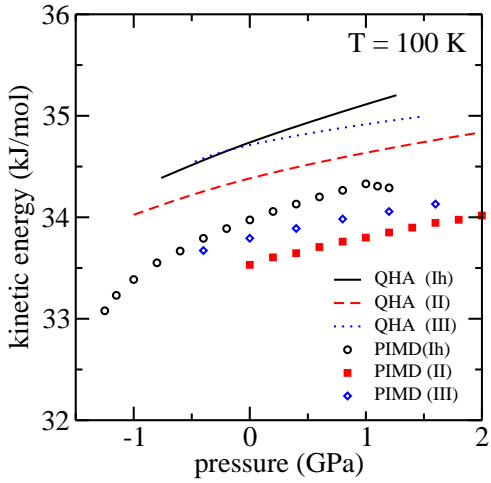}
\vspace{-0.9cm}
\caption{ Kinetic energy of the studied ice phases as a
function
of the pressure at 100 K. QHA results (lines) are compared to PIMD
simulations (symbols). Error bars of the simulation results are in
the order of the the symbol size.}
\label{fig:k_p}
\end{figure}

The pressure dependence of the kinetic energy at 100 K is displayed
in Fig. \ref{fig:k_p}. Here we also find that QHA results are shifted
with respect to the PIMD data. The shift remains nearly constant with
the pressure, except for the case of ice Ih at the highest simulated
pressures, where larger deviations are originated from the proximity
to the limits of the mechanical stability of this ice phase.

\subsection{Enthalpy}

\begin{figure}
\vspace{-0.3cm}
\includegraphics[width= 9cm]{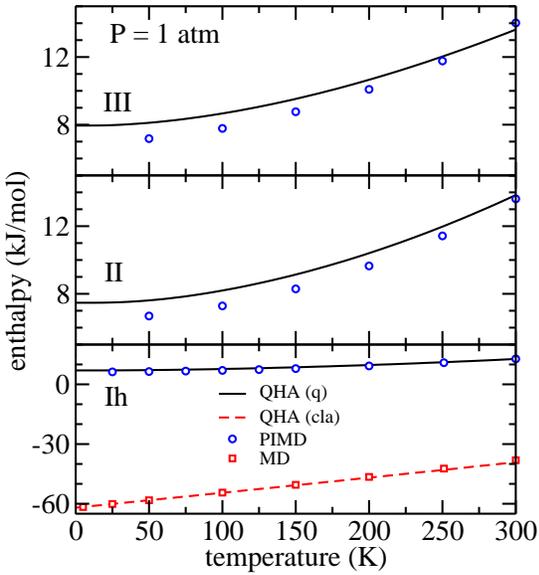}
\vspace{-0.9cm}
\caption{ Enthalpy of ice Ih, II, and III at atmospheric
pressure
calculated with the q-TIP4P/F water model. The QHA results (lines)
are compared to PIMD and MD simulations (symbols). For ice Ih both
quantum and classical limits are shown. Error bars of the simulation
results are less than the symbol size.}
\label{fig:h_t_pi}
\end{figure}

At atmospheric pressure the enthalpy of ice is nearly identical to
its internal energy as the term $PV$ results to be vanishingly small.
The enthalpy, $H$, of the three ice phases at atmospheric pressure
is presented in Fig. \ref{fig:h_t_pi} as a function of temperature.
For ice Ih we have compared both the classical and quantum limits
of the QHA to the corresponding simulation results. At the scale of
the figure the agreement is very good. The QHA zero-point energy of
ice Ih is 68.9 kJ/mol. The QHA systematically overestimates the enthalpy
of the three ice phases at low temperatures. In the case of ice Ih
we find that at 50 K the QHA result is 0.7 kJ/mol larger than the
PIMD result, while for ice II and III we obtain a larger deviation
of about 0.9 kJ/mol. Both QHA and PIMD results show that at 50 K the
enthalpy of the ice phases increases in the order Ih$<$II$<$III.
With respect to the enthalpy of ice II, the relative values found
for ice Ih and III at 50 K are -0.3 and 0.5 kJ/mol (PIMD) and -0.4
and 0.5 kJ/mol (QHA), respectively. Interestingly at higher temperature
the agreement between the PIMD and QHA data becomes better as a consequence
of an error compensation between kinetic and potential energy terms.
Thus the error of the QHA enthalpy estimation is lower than that found
for the kinetic energy in the previous Subsec. \ref{sub:Kinetic-energy}.

\begin{figure}
\vspace{-1.6cm}
\includegraphics[width= 9cm]{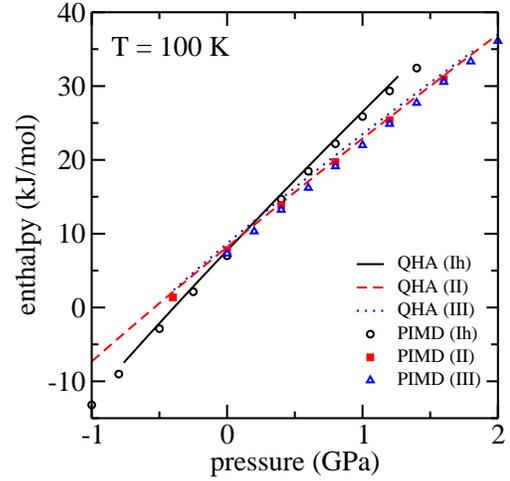}
\vspace{-0.9cm}
\caption{ Enthalpy of the studied ice phases as a function
of
the pressure at 100 K. QHA results (lines) are compared to PIMD
simulations
(symbols). Error bars of the simulation results are smaller than the
symbol size.}
\label{fig:h_p_pi}
\end{figure}

The pressure dependence of the enthalpy at 100 K is represented in
Fig. \ref{fig:h_p_pi}. The term $PV$ plays here an important role.
We have already seen in Fig. \ref{fig:h_t_pi} that at 100 K and atmospheric
pressure the QHA overestimates the reference enthalpy of PIMD simulations.
This behavior results nearly independent of the external pressure.
In Fig. \ref{fig:h_p_pi} the QHA result for the three ice phases
lies systematically slightly above the corresponding simulation results,
but the overall agreement can be considered as satisfactory in the
whole studied pressure range.

\subsection{Heat capacity }

\begin{figure}
\vspace{-0.8cm}
\includegraphics[width=8.5cm]{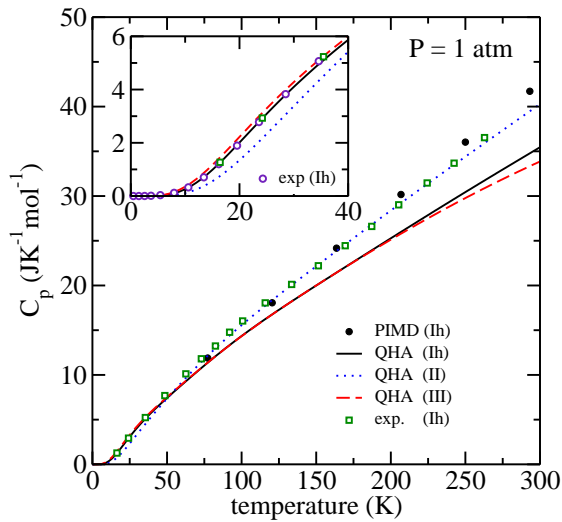}
\vspace{-0.5cm}
\caption{ Heat capacity of ice Ih, II, and III at
atmospheric
pressure. Lines are results derived by the QHA. Filled circles
correspond
to PIMD simulations of ice Ih. Experimental data for ice Ih are shown
by open circles (Ref. \onlinecite{smith07}) and open squares (Ref.
\onlinecite{giauque36}). The inset emphasizes the low temperature
limit.}
\label{fig:cp_t_exp}
\end{figure}

The heat capacity at constant pressure is defined as
\begin{equation}
C_{p}=\left(\frac{\partial H}{\partial T}\right)_{P}\quad.
\end{equation}
 The PIMD estimation of $C_{p}$ of ice Ih at atmospheric pressure
has been obtained from a numerical fit of the $H(T)$ curve shown
in Fig. \ref{fig:h_t_pi}. The temperature derivative was calculated
from a 4th degree polynomial fit of $H(T)$ in the interval 80-190
K. The result is plotted as filled circles in Fig. \ref{fig:cp_t_exp}.
For comparison the experimental $C_{p}$ data of ice Ih of Refs. \onlinecite{giauque36,smith07}
have been plotted as open circles and squares, respectively. We find
good agreement between the PIMD results and the experimental data.
The figure also includes lines showing the $C_{p}$ values derived
from the QHA of ice Ih, II, and III, with an inset that highlights
their low temperature behavior. Note that the QHA underestimates the
PIMD data of ice Ih at temperatures above 100 K. However, the QHA
result appears close to the simulation at 80 K. At low temperatures
(below 40 K, see the inset of Fig. \ref{fig:cp_t_exp}) the QHA for
ice Ih shows an excellent agreement to the experimental data. We expect
that such a good agreement will occur also in PIMD simulations. However,
the computational cost of these PIMD simulations increases as $1/T$,
along with the increase in the number of beads, so that a reliable
simulation of $C_{p}$ below 40 K becomes computationally too expensive.
Therefore the QHA seems to be a practical alternative to PIMD simulations
in the study of the low temperature heat capacity of ices.

The heat capacities of the three ice phases are significantly different
in the studied temperature range. In particular, below 50 K ice II
displays the lowest heat capacity at the studied pressure, while the
opposite behavior is observed above this temperature.

\begin{figure}
\vspace{-1.5cm}
\includegraphics[width= 9cm]{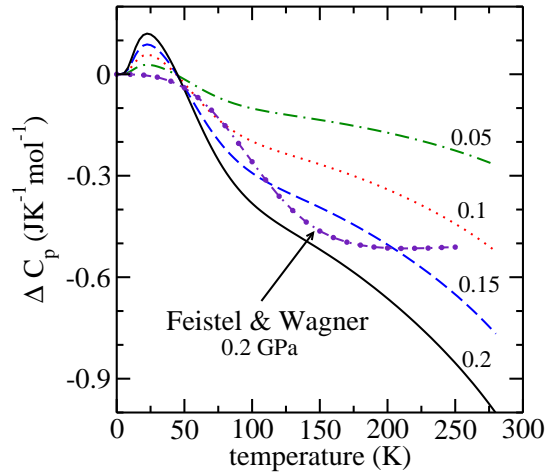}
\vspace{-0.8cm}
\caption{ Relative heat capacity,
$C_{p}(P)-C_{p}(P_{ref}=0)$,
of ice Ih as a function of temperature. The QHA results are shown
by lines labeled by the calculation pressure (in GPa). The curve with
symbols was derived by the empirical state equation of Ref.
\onlinecite{feistel06}
at a pressure of 0.2 GPa.}
\label{fig:relative_cp_t}
\end{figure}

It has been previously stressed that the pressure dependence of $C_{p}$
in ice Ih has not been studied experimentally. Thus the empirical
equation of state of Feistel and Wagner has been applied to extrapolate
the temperature dependence of $C_{p}$ at pressures up to 0.2 GPa.\citep{feistel06}
In particular the temperature dependence of the relative heat capacity
\begin{equation}
\Delta C_{p}=C_{p}(P)-C_{p}(P_{ref}=0)
\end{equation}
 derived at $P$=0.2 GPa with this empirical equation of state is
represented by full symbols in Fig. \ref{fig:relative_cp_t}. This
extrapolation of $\Delta C_{p}$ is characterized by having a rather
small slope at low and at high temperatures. The same qualitative
behavior is found for lower pressures.\citep{feistel06} On the contrary,
the QHA predicts a different overall behavior, with a maximum of $\Delta C_{p}$
at low temperatures ($\sim25$ K) and a conspicuous negative slope
at high temperatures. We believe that the QHA provides a more realistic
estimation of $\Delta C_{p}$ than that obtained by extrapolation
of the empirical equation of state. Our main argument is that the
QHA represents a realistic physical model of the ice phase.

\section{Conclusions}

In this work we have undertaken a systematic study of the capability
of the QHA to reproduce several thermodynamic properties of ice Ih,
II, and III as a function of both pressure and temperature. The studied
pressure range goes from -1 to 2 GPa, while the temperatures were
studied up to 300 K. Thus the region in the $P-T$ plane where the
QHA has been checked is much larger than the area where the studied
ice phases are found to be thermodynamically stable. Therefore an
important consideration of the present study is its generality, in
the sense that it is not limited to a small number of state points. 

The validity of the QHA is restricted by the presence of anharmonic
effects not included in the approximation. Thus a direct check of
the QHA is the comparison to numerical simulations that fully consider
the anharmonicity of the interatomic interactions. We have employed
the empirical q-TIP4P/F model for this comparison. Our expectation
that the validity of the QHA is largely independent of the employed
water model is based on the assumption that the anharmonicity of the
q-TIP4P/F potential is reasonably realistic in view of the large body
of experimental data reproduced by this model.\cite{habershon09,ramirez10,herrero11b}

The comparison of the QHA to the PIMD simulations shows a remarkable
overall agreement for the three ice phases. The best agreement has
been found for ice II, which is an ordered phase of ice. Both crystal
volume and enthalpy have been shown to be reasonably reproduced by
the QHA. This fact let us expect that the QHA could be appropriate
to study phase coexistence of ice phases as the slopes of phase coexistence
lines are a function of the differences in volume and enthalpy of
the corresponding ice phases (Clausius-Clapeyron relation). The QHA
is also particularly appropriate to study the low temperature limit
of certain thermodynamic properties as the heat capacity, bulk modulus
and internal energies. The computational cost of PI simulations increases
as $1/T$, so that the simulation of low temperature limits can be
prohibitively expensive by this approach. Further advantages offered
by the QHA are the lack of statistical errors and the easier checking
and correction of finite size errors. These advantages apply not only
compared to PI simulations, but even compared to classical MD.

The experimental heat capacity of ice Ih at low temperatures ($T<$40K)
is reproduced with remarkable accuracy with the QHA at atmospheric
pressure. Our results lead us to expect that this good agreement should
not deteriorate at higher pressures. We have shown that the temperature
dependence of heat capacities at finite pressures predicted by the
QHA differs in important aspects from a previous estimation based
on the extrapolation of the Feistel-Wagner equation of state for ice
Ih.\citealp{feistel06} Although we have not found experimental data
of the heat capacity of ice Ih above the normal pressure, we expect
that the temperature dependence predicted by the QHA is physically
sound because it is based on a reasonable physical model of ice.

An interesting aspect of the QHA is that it is sensible enough to
predict different anharmonic behaviors as a function of the employed
water model. The thermal expansion of ice Ih at low temperatures is
predicted by the QHA to be negative for the q-TIP4P/F and the TIP4P
potentials but positive (or slightly negative) for the TIP5P and ST2
models.\cite{koyama04} Moreover, the isotope effect in the crystal
volume of ice Ih is predicted by the QHA to be anomalous (as in the
experiment) with a DFT functional, but normal with the q-TIP4P/F model.
We stress that these differences in the QHA predictions are in agreement
to the results of available computer simulations.\cite{pamuk12}

At last we mention several lines where the QHA is likely to provide
useful results in relation to ice phases investigations: $i)$ the
dependence of thermodynamic variables with system size; $ii)$ low
temperature studies of ice phases where the computational cost of
PI simulations becomes increasingly large; $iii)$ isotope and quantum
effects in the phase diagram of ice; $iv)$ dependence of thermodynamic
variables on hydrogen disorder; $\mathit{v})$ check of improved water
models.

\acknowledgments
This work was supported by Ministerio de Ciencia e Innovación (Spain)
through Grant No. FIS2009-12721-C04-04, by Comunidad Autónoma de Madrid
through project MODELICO-CM/S2009ESP-1691, an at Stony Brook by DOE
award numbers DE-FG02-09ER16052 and DE-SC0003871 (MVFS).

\bibliographystyle{apsrev}

\begin{thebibliography}{70}
\expandafter\ifx\csname natexlab\endcsname\relax\def\natexlab#1{#1}\fi
\expandafter\ifx\csname bibnamefont\endcsname\relax
  \def\bibnamefont#1{#1}\fi
\expandafter\ifx\csname bibfnamefont\endcsname\relax
  \def\bibfnamefont#1{#1}\fi
\expandafter\ifx\csname citenamefont\endcsname\relax
  \def\citenamefont#1{#1}\fi
\expandafter\ifx\csname url\endcsname\relax
  \def\url#1{\texttt{#1}}\fi
\expandafter\ifx\csname urlprefix\endcsname\relax\def\urlprefix{URL }\fi
\providecommand{\bibinfo}[2]{#2}
\providecommand{\eprint}[2][]{\url{#2}}

\bibitem[{\citenamefont{Barker and Watts}(1969)}]{barker69}
\bibinfo{author}{\bibfnamefont{J.~A.} \bibnamefont{Barker}} \bibnamefont{and}
  \bibinfo{author}{\bibfnamefont{R.~O.} \bibnamefont{Watts}},
  \bibinfo{journal}{Chem. Phys. Lett.} \textbf{\bibinfo{volume}{3}},
  \bibinfo{pages}{144} (\bibinfo{year}{1969}).

\bibitem[{\citenamefont{Rahman and Stillinger}(1971)}]{rahman71}
\bibinfo{author}{\bibfnamefont{A.}~\bibnamefont{Rahman}} \bibnamefont{and}
  \bibinfo{author}{\bibfnamefont{F.~H.} \bibnamefont{Stillinger}},
  \bibinfo{journal}{J. Chem. Phys.} \textbf{\bibinfo{volume}{55}},
  \bibinfo{pages}{3336} (\bibinfo{year}{1971}).

\bibitem[{\citenamefont{Moore and Valeria}(2011)}]{moore11}
\bibinfo{author}{\bibfnamefont{E.~B.} \bibnamefont{Moore}} \bibnamefont{and}
  \bibinfo{author}{\bibfnamefont{M.}~\bibnamefont{Valeria}},
  \bibinfo{journal}{Nature} \textbf{\bibinfo{volume}{479}},
  \bibinfo{pages}{506} (\bibinfo{year}{2011}).

\bibitem[{\citenamefont{Jorgensen and Tirado-Rives}(2005)}]{jorgensen05}
\bibinfo{author}{\bibfnamefont{W.~L.} \bibnamefont{Jorgensen}}
  \bibnamefont{and}
  \bibinfo{author}{\bibfnamefont{J.}~\bibnamefont{Tirado-Rives}},
  \bibinfo{journal}{PNAS} \textbf{\bibinfo{volume}{102}}, \bibinfo{pages}{6665}
  (\bibinfo{year}{2005}).

\bibitem[{\citenamefont{Vega and Abascal}(2011)}]{vega11}
\bibinfo{author}{\bibfnamefont{C.}~\bibnamefont{Vega}} \bibnamefont{and}
  \bibinfo{author}{\bibfnamefont{J.~L.~F.} \bibnamefont{Abascal}},
  \bibinfo{journal}{Phys. Chem. Chem. Phys.} \textbf{\bibinfo{volume}{13}},
  \bibinfo{pages}{19663} (\bibinfo{year}{2011}).

\bibitem[{\citenamefont{Paesani et~al.}(2006)\citenamefont{Paesani, Zhang,
  Case, Cheatham, III, and Voth}}]{paesani06}
\bibinfo{author}{\bibfnamefont{F.}~\bibnamefont{Paesani}},
  \bibinfo{author}{\bibfnamefont{W.}~\bibnamefont{Zhang}},
  \bibinfo{author}{\bibfnamefont{D.~A.} \bibnamefont{Case}},
  \bibinfo{author}{\bibfnamefont{T.~E.} \bibnamefont{Cheatham}},
  \bibinfo{author}{\bibnamefont{III}}, \bibnamefont{and}
  \bibinfo{author}{\bibfnamefont{G.~A.} \bibnamefont{Voth}},
  \bibinfo{journal}{J. Chem. Phys.} \textbf{\bibinfo{volume}{125}},
  \bibinfo{eid}{184507} (\bibinfo{year}{2006}).

\bibitem[{\citenamefont{Habershon et~al.}(2009)\citenamefont{Habershon,
  Markland, and Manolopoulos}}]{habershon09}
\bibinfo{author}{\bibfnamefont{S.}~\bibnamefont{Habershon}},
  \bibinfo{author}{\bibfnamefont{T.~E.} \bibnamefont{Markland}},
  \bibnamefont{and} \bibinfo{author}{\bibfnamefont{D.~E.}
  \bibnamefont{Manolopoulos}}, \bibinfo{journal}{J. Chem. Phys.}
  \textbf{\bibinfo{volume}{131}}, \bibinfo{eid}{024501} (\bibinfo{year}{2009}).

\bibitem[{\citenamefont{Stern and Berne}(2001)}]{stern01}
\bibinfo{author}{\bibfnamefont{H.~A.} \bibnamefont{Stern}} \bibnamefont{and}
  \bibinfo{author}{\bibfnamefont{B.~J.} \bibnamefont{Berne}},
  \bibinfo{journal}{J. Chem. Phys.} \textbf{\bibinfo{volume}{115}},
  \bibinfo{pages}{7622} (\bibinfo{year}{2001}).

\bibitem[{\citenamefont{Fanourgakis and Xantheas}(2008)}]{fanourgakis08}
\bibinfo{author}{\bibfnamefont{G.~S.} \bibnamefont{Fanourgakis}}
  \bibnamefont{and} \bibinfo{author}{\bibfnamefont{S.~S.}
  \bibnamefont{Xantheas}}, \bibinfo{journal}{J. Chem. Phys.}
  \textbf{\bibinfo{volume}{128}}, \bibinfo{eid}{074506} (\bibinfo{year}{2008}).

\bibitem[{\citenamefont{Benoit et~al.}(2002)\citenamefont{Benoit, Romero, and
  Marx}}]{benoit02}
\bibinfo{author}{\bibfnamefont{M.}~\bibnamefont{Benoit}},
  \bibinfo{author}{\bibfnamefont{A.~H.} \bibnamefont{Romero}},
  \bibnamefont{and} \bibinfo{author}{\bibfnamefont{D.}~\bibnamefont{Marx}},
  \bibinfo{journal}{Phys. Rev. Lett.} \textbf{\bibinfo{volume}{89}},
  \bibinfo{pages}{145501} (\bibinfo{year}{2002}).

\bibitem[{\citenamefont{Chen et~al.}(2003)\citenamefont{Chen, Ivanov, Klein,
  and Parrinello}}]{chen03}
\bibinfo{author}{\bibfnamefont{B.}~\bibnamefont{Chen}},
  \bibinfo{author}{\bibfnamefont{I.}~\bibnamefont{Ivanov}},
  \bibinfo{author}{\bibfnamefont{M.~L.} \bibnamefont{Klein}}, \bibnamefont{and}
  \bibinfo{author}{\bibfnamefont{M.}~\bibnamefont{Parrinello}},
  \bibinfo{journal}{Phys. Rev. Lett.} \textbf{\bibinfo{volume}{91}},
  \bibinfo{pages}{215503} (\bibinfo{year}{2003}).

\bibitem[{\citenamefont{Fern\'{a}ndez-Serra and Artacho}(2004)}]{marivi04}
\bibinfo{author}{\bibfnamefont{M.~V.} \bibnamefont{Fern\'{a}ndez-Serra}}
  \bibnamefont{and} \bibinfo{author}{\bibfnamefont{E.}~\bibnamefont{Artacho}},
  \bibinfo{journal}{J. Chem. Phys.} \textbf{\bibinfo{volume}{121}},
  \bibinfo{pages}{11136} (\bibinfo{year}{2004}).

\bibitem[{\citenamefont{Fern\'andez-Serra and Artacho}(2006)}]{marivi06}
\bibinfo{author}{\bibfnamefont{M.~V.} \bibnamefont{Fern\'andez-Serra}}
  \bibnamefont{and} \bibinfo{author}{\bibfnamefont{E.}~\bibnamefont{Artacho}},
  \bibinfo{journal}{Phys. Rev. Lett.} \textbf{\bibinfo{volume}{96}},
  \bibinfo{pages}{016404} (\bibinfo{year}{2006}).

\bibitem[{\citenamefont{Morrone and Car}(2008)}]{morrone08}
\bibinfo{author}{\bibfnamefont{J.~A.} \bibnamefont{Morrone}} \bibnamefont{and}
  \bibinfo{author}{\bibfnamefont{R.}~\bibnamefont{Car}},
  \bibinfo{journal}{Phys. Rev. Lett.} \textbf{\bibinfo{volume}{101}},
  \bibinfo{pages}{017801} (\bibinfo{year}{2008}).

\bibitem[{\citenamefont{Yoo et~al.}(2009)\citenamefont{Yoo, Zeng, and
  Xantheas}}]{yoo09}
\bibinfo{author}{\bibfnamefont{S.}~\bibnamefont{Yoo}},
  \bibinfo{author}{\bibfnamefont{X.~C.} \bibnamefont{Zeng}}, \bibnamefont{and}
  \bibinfo{author}{\bibfnamefont{S.~S.} \bibnamefont{Xantheas}},
  \bibinfo{journal}{J. Chem. Phys.} \textbf{\bibinfo{volume}{130}},
  \bibinfo{eid}{221102} (\bibinfo{year}{2009}).

\bibitem[{\citenamefont{Dion et~al.}(2004)\citenamefont{Dion, Rydberg,
  Schr\"oder, Langreth, and Lundqvist}}]{dion04}
\bibinfo{author}{\bibfnamefont{M.}~\bibnamefont{Dion}},
  \bibinfo{author}{\bibfnamefont{H.}~\bibnamefont{Rydberg}},
  \bibinfo{author}{\bibfnamefont{E.}~\bibnamefont{Schr\"oder}},
  \bibinfo{author}{\bibfnamefont{D.~C.} \bibnamefont{Langreth}},
  \bibnamefont{and} \bibinfo{author}{\bibfnamefont{B.~I.}
  \bibnamefont{Lundqvist}}, \bibinfo{journal}{Phys. Rev. Lett.}
  \textbf{\bibinfo{volume}{92}}, \bibinfo{pages}{246401}
  (\bibinfo{year}{2004}).

\bibitem[{\citenamefont{Murray and Galli}(2012)}]{murray11}
\bibinfo{author}{\bibfnamefont{E.~D.} \bibnamefont{Murray}} \bibnamefont{and}
  \bibinfo{author}{\bibfnamefont{G.}~\bibnamefont{Galli}},
  \bibinfo{journal}{Phys. Rev. Lett.} \textbf{\bibinfo{volume}{108}},
  \bibinfo{pages}{105502} (\bibinfo{year}{2012}).

\bibitem[{\citenamefont{Wang et~al.}(2011)\citenamefont{Wang,
  Rom\'{a}n-P\'{e}rez, Soler, Artacho, and Fern\'{a}ndez-Serra}}]{wang11}
\bibinfo{author}{\bibfnamefont{J.}~\bibnamefont{Wang}},
  \bibinfo{author}{\bibfnamefont{G.}~\bibnamefont{Rom\'{a}n-P\'{e}rez}},
  \bibinfo{author}{\bibfnamefont{J.~M.} \bibnamefont{Soler}},
  \bibinfo{author}{\bibfnamefont{E.}~\bibnamefont{Artacho}}, \bibnamefont{and}
  \bibinfo{author}{\bibfnamefont{M.-V.} \bibnamefont{Fern\'{a}ndez-Serra}},
  \bibinfo{journal}{J. Chem. Phys.} \textbf{\bibinfo{volume}{134}},
  \bibinfo{pages}{024516} (\bibinfo{year}{2011}).

\bibitem[{\citenamefont{Santra et~al.}(2011)\citenamefont{Santra,
  Klime\ifmmode~\check{s}\else \v{s}\fi{}, Alf\`e, Tkatchenko, Slater,
  Michaelides, Car, and Scheffler}}]{santra11}
\bibinfo{author}{\bibfnamefont{B.}~\bibnamefont{Santra}},
  \bibinfo{author}{\bibfnamefont{J.}~\bibnamefont{Klime\ifmmode~\check{s}\else
  \v{s}\fi{}}}, \bibinfo{author}{\bibfnamefont{D.}~\bibnamefont{Alf\`e}},
  \bibinfo{author}{\bibfnamefont{A.}~\bibnamefont{Tkatchenko}},
  \bibinfo{author}{\bibfnamefont{B.}~\bibnamefont{Slater}},
  \bibinfo{author}{\bibfnamefont{A.}~\bibnamefont{Michaelides}},
  \bibinfo{author}{\bibfnamefont{R.}~\bibnamefont{Car}}, \bibnamefont{and}
  \bibinfo{author}{\bibfnamefont{M.}~\bibnamefont{Scheffler}},
  \bibinfo{journal}{Phys. Rev. Lett.} \textbf{\bibinfo{volume}{107}},
  \bibinfo{pages}{185701} (\bibinfo{year}{2011}).

\bibitem[{\citenamefont{R{\"{o}}ttger et~al.}(2012)\citenamefont{R{\"{o}}ttger,
  Endriss, Ihringer, Doyle, and Kuhs}}]{rottger12}
\bibinfo{author}{\bibfnamefont{K.}~\bibnamefont{R{\"{o}}ttger}},
  \bibinfo{author}{\bibfnamefont{A.}~\bibnamefont{Endriss}},
  \bibinfo{author}{\bibfnamefont{J.}~\bibnamefont{Ihringer}},
  \bibinfo{author}{\bibfnamefont{S.}~\bibnamefont{Doyle}}, \bibnamefont{and}
  \bibinfo{author}{\bibfnamefont{W.~F.} \bibnamefont{Kuhs}},
  \bibinfo{journal}{Acta Cryst. B} \textbf{\bibinfo{volume}{68}},
  \bibinfo{pages}{91} (\bibinfo{year}{2012}).

\bibitem[{\citenamefont{R{\"{o}}ttger et~al.}(1994)\citenamefont{R{\"{o}}ttger,
  Endriss, Ihringer, Doyle, and Kuhs}}]{rottger94}
\bibinfo{author}{\bibfnamefont{K.}~\bibnamefont{R{\"{o}}ttger}},
  \bibinfo{author}{\bibfnamefont{A.}~\bibnamefont{Endriss}},
  \bibinfo{author}{\bibfnamefont{J.}~\bibnamefont{Ihringer}},
  \bibinfo{author}{\bibfnamefont{S.}~\bibnamefont{Doyle}}, \bibnamefont{and}
  \bibinfo{author}{\bibfnamefont{W.~F.} \bibnamefont{Kuhs}},
  \bibinfo{journal}{Acta Crys. B} \textbf{\bibinfo{volume}{50}},
  \bibinfo{pages}{644} (\bibinfo{year}{1994}).

\bibitem[{\citenamefont{Pamuk et~al.}(2012)\citenamefont{Pamuk, Soler,
  Ram\'{i}rez, Herrero, Stephens, Allen, and Fern\'{a}ndez-Serra}}]{pamuk12}
\bibinfo{author}{\bibfnamefont{B.}~\bibnamefont{Pamuk}},
  \bibinfo{author}{\bibfnamefont{J.~M.} \bibnamefont{Soler}},
  \bibinfo{author}{\bibfnamefont{R.}~\bibnamefont{Ram\'{i}rez}},
  \bibinfo{author}{\bibfnamefont{C.~P.} \bibnamefont{Herrero}},
  \bibinfo{author}{\bibfnamefont{P.~W.} \bibnamefont{Stephens}},
  \bibinfo{author}{\bibfnamefont{P.~B.} \bibnamefont{Allen}}, \bibnamefont{and}
  \bibinfo{author}{\bibfnamefont{M.-V.} \bibnamefont{Fern\'{a}ndez-Serra}},
  \bibinfo{journal}{Phys. Rev. Lett., in press.}  (\bibinfo{year}{2012}).

\bibitem[{\citenamefont{Kuharski and Rossky}(1985)}]{kuharski85}
\bibinfo{author}{\bibfnamefont{R.~A.} \bibnamefont{Kuharski}} \bibnamefont{and}
  \bibinfo{author}{\bibfnamefont{P.~J.} \bibnamefont{Rossky}},
  \bibinfo{journal}{J. Chem. Phys.} \textbf{\bibinfo{volume}{82}},
  \bibinfo{pages}{5164} (\bibinfo{year}{1985}).

\bibitem[{\citenamefont{Mahoney and Jorgensen}(2001)}]{mahoney01}
\bibinfo{author}{\bibfnamefont{M.~W.} \bibnamefont{Mahoney}} \bibnamefont{and}
  \bibinfo{author}{\bibfnamefont{W.~L.} \bibnamefont{Jorgensen}},
  \bibinfo{journal}{J. Chem. Phys.} \textbf{\bibinfo{volume}{115}},
  \bibinfo{pages}{10758} (\bibinfo{year}{2001}).

\bibitem[{\citenamefont{{Hern\'{a}ndez de la Pe\~{n}a} and
  Kusalik}(2004)}]{hernandezpena04}
\bibinfo{author}{\bibfnamefont{L.}~\bibnamefont{{Hern\'{a}ndez de la
  Pe\~{n}a}}} \bibnamefont{and} \bibinfo{author}{\bibfnamefont{P.~G.}
  \bibnamefont{Kusalik}}, \bibinfo{journal}{J. Chem. Phys.}
  \textbf{\bibinfo{volume}{121}}, \bibinfo{pages}{5992} (\bibinfo{year}{2004}).

\bibitem[{\citenamefont{Shiga and Shinoda}(2005)}]{shiga05}
\bibinfo{author}{\bibfnamefont{M.}~\bibnamefont{Shiga}} \bibnamefont{and}
  \bibinfo{author}{\bibfnamefont{W.}~\bibnamefont{Shinoda}},
  \bibinfo{journal}{J. Chem. Phys.} \textbf{\bibinfo{volume}{123}},
  \bibinfo{eid}{134502} (\bibinfo{year}{2005}).

\bibitem[{\citenamefont{Paesani et~al.}(2007)\citenamefont{Paesani, Iuchi, and
  Voth}}]{paesani07}
\bibinfo{author}{\bibfnamefont{F.}~\bibnamefont{Paesani}},
  \bibinfo{author}{\bibfnamefont{S.}~\bibnamefont{Iuchi}}, \bibnamefont{and}
  \bibinfo{author}{\bibfnamefont{G.~A.} \bibnamefont{Voth}},
  \bibinfo{journal}{J. Chem. Phys.} \textbf{\bibinfo{volume}{127}},
  \bibinfo{eid}{074506} (\bibinfo{year}{2007}).

\bibitem[{\citenamefont{McBride et~al.}(2009)\citenamefont{McBride, Vega, Noya,
  Ram\'{\i}rez, and Ses\'{e}}}]{mcbride09}
\bibinfo{author}{\bibfnamefont{C.}~\bibnamefont{McBride}},
  \bibinfo{author}{\bibfnamefont{C.}~\bibnamefont{Vega}},
  \bibinfo{author}{\bibfnamefont{E.~G.} \bibnamefont{Noya}},
  \bibinfo{author}{\bibfnamefont{R.}~\bibnamefont{Ram\'{\i}rez}},
  \bibnamefont{and} \bibinfo{author}{\bibfnamefont{L.~M.}
  \bibnamefont{Ses\'{e}}}, \bibinfo{journal}{J. Chem. Phys.}
  \textbf{\bibinfo{volume}{131}}, \bibinfo{eid}{024506} (\bibinfo{year}{2009}).

\bibitem[{\citenamefont{Vega et~al.}(2010)\citenamefont{Vega, Conde, McBride,
  Abascal, Noya, Ramirez, and Ses\'{e}}}]{vega10}
\bibinfo{author}{\bibfnamefont{C.}~\bibnamefont{Vega}},
  \bibinfo{author}{\bibfnamefont{M.~M.} \bibnamefont{Conde}},
  \bibinfo{author}{\bibfnamefont{C.}~\bibnamefont{McBride}},
  \bibinfo{author}{\bibfnamefont{J.~L.~F.} \bibnamefont{Abascal}},
  \bibinfo{author}{\bibfnamefont{E.~G.} \bibnamefont{Noya}},
  \bibinfo{author}{\bibfnamefont{R.}~\bibnamefont{Ramirez}}, \bibnamefont{and}
  \bibinfo{author}{\bibfnamefont{L.~M.} \bibnamefont{Ses\'{e}}},
  \bibinfo{journal}{J. Chem. Phys.} \textbf{\bibinfo{volume}{132}},
  \bibinfo{eid}{046101} (\bibinfo{year}{2010}).

\bibitem[{\citenamefont{Ram\'{\i}rez and Herrero}(2010)}]{ramirez10}
\bibinfo{author}{\bibfnamefont{R.}~\bibnamefont{Ram\'{\i}rez}}
  \bibnamefont{and} \bibinfo{author}{\bibfnamefont{C.~P.}
  \bibnamefont{Herrero}}, \bibinfo{journal}{J. Chem. Phys.}
  \textbf{\bibinfo{volume}{133}}, \bibinfo{eid}{144511} (\bibinfo{year}{2010}).

\bibitem[{\citenamefont{Herrero and
  Ram\'{\i}rez}(2011{\natexlab{a}})}]{herrero11}
\bibinfo{author}{\bibfnamefont{C.~P.} \bibnamefont{Herrero}} \bibnamefont{and}
  \bibinfo{author}{\bibfnamefont{R.}~\bibnamefont{Ram\'{\i}rez}},
  \bibinfo{journal}{J. Chem. Phys.} \textbf{\bibinfo{volume}{134}},
  \bibinfo{eid}{094510} (\bibinfo{year}{2011}{\natexlab{a}}).

\bibitem[{\citenamefont{Herrero and
  Ram\'{\i}rez}(2011{\natexlab{b}})}]{herrero11b}
\bibinfo{author}{\bibfnamefont{C.~P.} \bibnamefont{Herrero}} \bibnamefont{and}
  \bibinfo{author}{\bibfnamefont{R.}~\bibnamefont{Ram\'{\i}rez}},
  \bibinfo{journal}{Phys. Rev. B} \textbf{\bibinfo{volume}{84}},
  \bibinfo{pages}{224112} (\bibinfo{year}{2011}{\natexlab{b}}).

\bibitem[{\citenamefont{Ram\'{\i}rez and Herrero}(2011)}]{ramirez11}
\bibinfo{author}{\bibfnamefont{R.}~\bibnamefont{Ram\'{\i}rez}}
  \bibnamefont{and} \bibinfo{author}{\bibfnamefont{C.~P.}
  \bibnamefont{Herrero}}, \bibinfo{journal}{Phys. Rev. B}
  \textbf{\bibinfo{volume}{84}}, \bibinfo{pages}{064130}
  (\bibinfo{year}{2011}).

\bibitem[{\citenamefont{Sanz et~al.}(2004)\citenamefont{Sanz, Vega, Abascal,
  and MacDowell}}]{sanz04}
\bibinfo{author}{\bibfnamefont{E.}~\bibnamefont{Sanz}},
  \bibinfo{author}{\bibfnamefont{C.}~\bibnamefont{Vega}},
  \bibinfo{author}{\bibfnamefont{J.~L.~F.} \bibnamefont{Abascal}},
  \bibnamefont{and} \bibinfo{author}{\bibfnamefont{L.~G.}
  \bibnamefont{MacDowell}}, \bibinfo{journal}{Phys. Rev. Lett.}
  \textbf{\bibinfo{volume}{92}}, \bibinfo{pages}{255701}
  (\bibinfo{year}{2004}).

\bibitem[{\citenamefont{Vega et~al.}(2008)\citenamefont{Vega, Sanz, Abascal,
  and Noya}}]{vega09}
\bibinfo{author}{\bibfnamefont{C.}~\bibnamefont{Vega}},
  \bibinfo{author}{\bibfnamefont{E.}~\bibnamefont{Sanz}},
  \bibinfo{author}{\bibfnamefont{J.~L.~F.} \bibnamefont{Abascal}},
  \bibnamefont{and} \bibinfo{author}{\bibfnamefont{E.~G.} \bibnamefont{Noya}},
  \bibinfo{journal}{J. Phys.: Condens. Matter} \textbf{\bibinfo{volume}{20}},
  \bibinfo{pages}{153101} (\bibinfo{year}{2008}).

\bibitem[{\citenamefont{Markland and Manolopoulos}(2008)}]{markland08}
\bibinfo{author}{\bibfnamefont{T.~E.} \bibnamefont{Markland}} \bibnamefont{and}
  \bibinfo{author}{\bibfnamefont{D.~E.} \bibnamefont{Manolopoulos}},
  \bibinfo{journal}{Chem. Phys. Lett.} \textbf{\bibinfo{volume}{464}},
  \bibinfo{pages}{256 } (\bibinfo{year}{2008}).

\bibitem[{\citenamefont{Ceriotti et~al.}(2011)\citenamefont{Ceriotti,
  Manolopoulos, and Parrinello}}]{ceriotti11}
\bibinfo{author}{\bibfnamefont{M.}~\bibnamefont{Ceriotti}},
  \bibinfo{author}{\bibfnamefont{D.~E.} \bibnamefont{Manolopoulos}},
  \bibnamefont{and}
  \bibinfo{author}{\bibfnamefont{M.}~\bibnamefont{Parrinello}},
  \bibinfo{journal}{J. Chem. Phys.} \textbf{\bibinfo{volume}{134}},
  \bibinfo{eid}{084104} (\bibinfo{year}{2011}).

\bibitem[{\citenamefont{Tanaka}(2001)}]{tanaka01}
\bibinfo{author}{\bibfnamefont{H.}~\bibnamefont{Tanaka}},
  \bibinfo{journal}{Journal of Molecular Liquids}
  \textbf{\bibinfo{volume}{90}}, \bibinfo{pages}{323 } (\bibinfo{year}{2001}).

\bibitem[{\citenamefont{Tse et~al.}(1999{\natexlab{a}})\citenamefont{Tse,
  Shpakov, and Belosludov}}]{tse99}
\bibinfo{author}{\bibfnamefont{J.~S.} \bibnamefont{Tse}},
  \bibinfo{author}{\bibfnamefont{V.~P.} \bibnamefont{Shpakov}},
  \bibnamefont{and} \bibinfo{author}{\bibfnamefont{V.~R.}
  \bibnamefont{Belosludov}}, \bibinfo{journal}{J. Chem. Phys.}
  \textbf{\bibinfo{volume}{111}}, \bibinfo{pages}{11111}
  (\bibinfo{year}{1999}{\natexlab{a}}).

\bibitem[{\citenamefont{Tse et~al.}(1999{\natexlab{b}})\citenamefont{Tse, Klug,
  Tulk, Swainson, Svensson, Loong, Shpakov, Belosludov, Belosludov, and
  Kawazoe}}]{tse99b}
\bibinfo{author}{\bibfnamefont{J.}~\bibnamefont{Tse}},
  \bibinfo{author}{\bibfnamefont{D.~D.} \bibnamefont{Klug}},
  \bibinfo{author}{\bibfnamefont{C.~A.} \bibnamefont{Tulk}},
  \bibinfo{author}{\bibfnamefont{I.~P.} \bibnamefont{Swainson}},
  \bibinfo{author}{\bibfnamefont{E.~C.} \bibnamefont{Svensson}},
  \bibinfo{author}{\bibfnamefont{C.-K.} \bibnamefont{Loong}},
  \bibinfo{author}{\bibfnamefont{V.~P.} \bibnamefont{Shpakov}},
  \bibinfo{author}{\bibfnamefont{V.~R.} \bibnamefont{Belosludov}},
  \bibinfo{author}{\bibfnamefont{R.~V.} \bibnamefont{Belosludov}},
  \bibnamefont{and} \bibinfo{author}{\bibfnamefont{Y.}~\bibnamefont{Kawazoe}},
  \bibinfo{journal}{Nature} \textbf{\bibinfo{volume}{400}},
  \bibinfo{pages}{647} (\bibinfo{year}{1999}{\natexlab{b}}).

\bibitem[{\citenamefont{Pauling}(1935)}]{pauling35}
\bibinfo{author}{\bibfnamefont{L.}~\bibnamefont{Pauling}}, \bibinfo{journal}{J.
  Am. Chem. Soc.} \textbf{\bibinfo{volume}{57}}, \bibinfo{pages}{2680}
  (\bibinfo{year}{1935}).

\bibitem[{\citenamefont{Lobban et~al.}(2000)\citenamefont{Lobban, Finney, and
  Kuhs}}]{lobban00}
\bibinfo{author}{\bibfnamefont{C.}~\bibnamefont{Lobban}},
  \bibinfo{author}{\bibfnamefont{J.~L.} \bibnamefont{Finney}},
  \bibnamefont{and} \bibinfo{author}{\bibfnamefont{W.~F.} \bibnamefont{Kuhs}},
  \bibinfo{journal}{J. Chem. Phys} \textbf{\bibinfo{volume}{112}},
  \bibinfo{pages}{7169} (\bibinfo{year}{2000}).

\bibitem[{\citenamefont{Buch et~al.}(1998)\citenamefont{Buch, Sandler, and
  Sadlej}}]{buch98}
\bibinfo{author}{\bibfnamefont{V.}~\bibnamefont{Buch}},
  \bibinfo{author}{\bibfnamefont{P.}~\bibnamefont{Sandler}}, \bibnamefont{and}
  \bibinfo{author}{\bibfnamefont{J.}~\bibnamefont{Sadlej}},
  \bibinfo{journal}{J. Chem. Phys} \textbf{\bibinfo{volume}{102}},
  \bibinfo{pages}{8641} (\bibinfo{year}{1998}).

\bibitem[{\citenamefont{Habershon and Manolopoulos}(2011)}]{haberschon11}
\bibinfo{author}{\bibfnamefont{S.}~\bibnamefont{Habershon}} \bibnamefont{and}
  \bibinfo{author}{\bibfnamefont{D.~E.} \bibnamefont{Manolopoulos}},
  \bibinfo{journal}{Phys. Chem. Chem. Phys.} \textbf{\bibinfo{volume}{13}},
  \bibinfo{pages}{19714} (\bibinfo{year}{2011}).

\bibitem[{\citenamefont{Cogoni et~al.}(2011)\citenamefont{Cogoni, D'Aguanno,
  Kuleshova, and Hofmann}}]{cogoni11}
\bibinfo{author}{\bibfnamefont{M.}~\bibnamefont{Cogoni}},
  \bibinfo{author}{\bibfnamefont{B.}~\bibnamefont{D'Aguanno}},
  \bibinfo{author}{\bibfnamefont{L.~N.} \bibnamefont{Kuleshova}},
  \bibnamefont{and} \bibinfo{author}{\bibfnamefont{D.~W.~M.}
  \bibnamefont{Hofmann}}, \bibinfo{journal}{J. Chem. Phys.}
  \textbf{\bibinfo{volume}{134}}, \bibinfo{eid}{204506} (\bibinfo{year}{2011}).

\bibitem[{\citenamefont{Hayward and Reimers}(1997)}]{hayward87}
\bibinfo{author}{\bibfnamefont{J.~A.} \bibnamefont{Hayward}} \bibnamefont{and}
  \bibinfo{author}{\bibfnamefont{J.~R.} \bibnamefont{Reimers}},
  \bibinfo{journal}{J. Chem. Phys} \textbf{\bibinfo{volume}{106}},
  \bibinfo{pages}{1518} (\bibinfo{year}{1997}).

\bibitem[{\citenamefont{Kamb et~al.}(1971)\citenamefont{Kamb, Hamilton,
  LaPlaca, and Prakash}}]{kamb71}
\bibinfo{author}{\bibfnamefont{B.}~\bibnamefont{Kamb}},
  \bibinfo{author}{\bibfnamefont{W.~C.} \bibnamefont{Hamilton}},
  \bibinfo{author}{\bibfnamefont{S.~J.} \bibnamefont{LaPlaca}},
  \bibnamefont{and} \bibinfo{author}{\bibfnamefont{A.}~\bibnamefont{Prakash}},
  \bibinfo{journal}{J. Chem. Phys} \textbf{\bibinfo{volume}{55}},
  \bibinfo{pages}{1934} (\bibinfo{year}{1971}).

\bibitem[{\citenamefont{Feynman}(1972)}]{feynman72}
\bibinfo{author}{\bibfnamefont{R.~P.} \bibnamefont{Feynman}},
  \emph{\bibinfo{title}{Statistical Mechanics}}
  (\bibinfo{publisher}{Addison-Wesley}, \bibinfo{address}{New York},
  \bibinfo{year}{1972}).

\bibitem[{\citenamefont{Gillan}(1988)}]{gillan88}
\bibinfo{author}{\bibfnamefont{M.~J.} \bibnamefont{Gillan}},
  \bibinfo{journal}{Phil. Mag. A} \textbf{\bibinfo{volume}{58}},
  \bibinfo{pages}{257} (\bibinfo{year}{1988}).

\bibitem[{\citenamefont{Ceperley}(1995)}]{ceperley95}
\bibinfo{author}{\bibfnamefont{D.~M.} \bibnamefont{Ceperley}},
  \bibinfo{journal}{Rev. Mod. Phys.} \textbf{\bibinfo{volume}{67}},
  \bibinfo{pages}{279} (\bibinfo{year}{1995}).

\bibitem[{\citenamefont{Chakravarty}(1997)}]{chakravarty97}
\bibinfo{author}{\bibfnamefont{C.}~\bibnamefont{Chakravarty}},
  \bibinfo{journal}{Int. Rev. Phys. Chem.} \textbf{\bibinfo{volume}{16}},
  \bibinfo{pages}{421} (\bibinfo{year}{1997}).

\bibitem[{\citenamefont{Martyna et~al.}(1999)\citenamefont{Martyna, Hughes, and
  Tuckerman}}]{ma99}
\bibinfo{author}{\bibfnamefont{G.~J.} \bibnamefont{Martyna}},
  \bibinfo{author}{\bibfnamefont{A.}~\bibnamefont{Hughes}}, \bibnamefont{and}
  \bibinfo{author}{\bibfnamefont{M.~E.} \bibnamefont{Tuckerman}},
  \bibinfo{journal}{J. Chem. Phys.} \textbf{\bibinfo{volume}{110}},
  \bibinfo{pages}{3275} (\bibinfo{year}{1999}).

\bibitem[{\citenamefont{Tuckerman}(2002)}]{tu02}
\bibinfo{author}{\bibfnamefont{M.~E.} \bibnamefont{Tuckerman}}, in
  \emph{\bibinfo{booktitle}{Quantum Simulations of Complex Many--Body Systems:
  From Theory to Algorithms}}, edited by
  \bibinfo{editor}{\bibfnamefont{J.}~\bibnamefont{Grotendorst}},
  \bibinfo{editor}{\bibfnamefont{D.}~\bibnamefont{Marx}}, \bibnamefont{and}
  \bibinfo{editor}{\bibfnamefont{A.}~\bibnamefont{Muramatsu}}
  (\bibinfo{publisher}{NIC}, \bibinfo{address}{FZ J\"ulich},
  \bibinfo{year}{2002}), p. \bibinfo{pages}{269}.

\bibitem[{\citenamefont{Tuckerman and Hughes}(1998)}]{tu98}
\bibinfo{author}{\bibfnamefont{M.~E.} \bibnamefont{Tuckerman}}
  \bibnamefont{and} \bibinfo{author}{\bibfnamefont{A.}~\bibnamefont{Hughes}},
  in \emph{\bibinfo{booktitle}{Classical \& Quantum Dynamics in Condensed Phase
  Simulations}}, edited by \bibinfo{editor}{\bibfnamefont{B.~J.}
  \bibnamefont{Berne}} \bibnamefont{and} \bibinfo{editor}{\bibfnamefont{D.~F.}
  \bibnamefont{Coker}} (\bibinfo{publisher}{Word Scientific},
  \bibinfo{address}{Singapore}, \bibinfo{year}{1998}), p. \bibinfo{pages}{311}.

\bibitem[{\citenamefont{Tuckerman et~al.}(1993)\citenamefont{Tuckerman, Berne,
  Martyna, and Klein}}]{tuckerman93}
\bibinfo{author}{\bibfnamefont{M.~E.} \bibnamefont{Tuckerman}},
  \bibinfo{author}{\bibfnamefont{B.~J.} \bibnamefont{Berne}},
  \bibinfo{author}{\bibfnamefont{G.~J.} \bibnamefont{Martyna}},
  \bibnamefont{and} \bibinfo{author}{\bibfnamefont{M.~L.} \bibnamefont{Klein}},
  \bibinfo{journal}{J. Chem. Phys.} \textbf{\bibinfo{volume}{99}},
  \bibinfo{pages}{2796} (\bibinfo{year}{1993}).

\bibitem[{\citenamefont{Martyna et~al.}(1996)\citenamefont{Martyna, Tuckerman,
  Tobias, and Klein}}]{ma96}
\bibinfo{author}{\bibfnamefont{G.~J.} \bibnamefont{Martyna}},
  \bibinfo{author}{\bibfnamefont{M.~E.} \bibnamefont{Tuckerman}},
  \bibinfo{author}{\bibfnamefont{D.~J.} \bibnamefont{Tobias}},
  \bibnamefont{and} \bibinfo{author}{\bibfnamefont{M.~L.} \bibnamefont{Klein}},
  \bibinfo{journal}{Mol. Phys.} \textbf{\bibinfo{volume}{87}},
  \bibinfo{pages}{1117} (\bibinfo{year}{1996}).

\bibitem[{\citenamefont{Pacheco}(1997)}]{pacheco97}
\bibinfo{author}{\bibfnamefont{P.}~\bibnamefont{Pacheco}},
  \emph{\bibinfo{title}{Parallel Programming with MPI}}
  (\bibinfo{publisher}{Morgan-Kaufmann}, \bibinfo{address}{San Francisco},
  \bibinfo{year}{1997}).

\bibitem[{\citenamefont{Kresse et~al.}(1995)\citenamefont{Kresse,
  Furthm\"uller, and Hafner}}]{kresse95}
\bibinfo{author}{\bibfnamefont{G.}~\bibnamefont{Kresse}},
  \bibinfo{author}{\bibfnamefont{J.}~\bibnamefont{Furthm\"uller}},
  \bibnamefont{and} \bibinfo{author}{\bibfnamefont{J.}~\bibnamefont{Hafner}},
  \bibinfo{journal}{Europhys. Lett.} \textbf{\bibinfo{volume}{32}},
  \bibinfo{pages}{729} (\bibinfo{year}{1995}).

\bibitem[{\citenamefont{Alf\`e et~al.}(2001)\citenamefont{Alf\`e, Price, and
  Gillan}}]{alfe01}
\bibinfo{author}{\bibfnamefont{D.}~\bibnamefont{Alf\`e}},
  \bibinfo{author}{\bibfnamefont{G.~D.} \bibnamefont{Price}}, \bibnamefont{and}
  \bibinfo{author}{\bibfnamefont{M.~J.} \bibnamefont{Gillan}},
  \bibinfo{journal}{Phys. Rev. B} \textbf{\bibinfo{volume}{64}},
  \bibinfo{pages}{045123} (\bibinfo{year}{2001}).

\bibitem[{\citenamefont{Venkataraman and Sahni}(1970)}]{venkataraman70}
\bibinfo{author}{\bibfnamefont{G.}~\bibnamefont{Venkataraman}}
  \bibnamefont{and} \bibinfo{author}{\bibfnamefont{V.~C.} \bibnamefont{Sahni}},
  \bibinfo{journal}{Rev. Mod. Phys.} \textbf{\bibinfo{volume}{42}},
  \bibinfo{pages}{409} (\bibinfo{year}{1970}).

\bibitem[{\citenamefont{Ashcroft and Mermin}(1976)}]{ashcroft76}
\bibinfo{author}{\bibfnamefont{N.~W.} \bibnamefont{Ashcroft}} \bibnamefont{and}
  \bibinfo{author}{\bibfnamefont{D.~N.} \bibnamefont{Mermin}},
  \emph{\bibinfo{title}{{Solid State Physics}}} (\bibinfo{publisher}{Saunders
  College}, \bibinfo{address}{Philadelphia}, \bibinfo{year}{1976}).

\bibitem[{\citenamefont{Libowitzky}(1999)}]{libowitzky99}
\bibinfo{author}{\bibfnamefont{E.}~\bibnamefont{Libowitzky}},
  \bibinfo{journal}{Monatsh. Chem.} \textbf{\bibinfo{volume}{130}},
  \bibinfo{pages}{1047} (\bibinfo{year}{1999}).

\bibitem[{\citenamefont{Bratos et~al.}(2009)\citenamefont{Bratos, Leicknam, and
  Pommeret}}]{bratos09}
\bibinfo{author}{\bibfnamefont{S.}~\bibnamefont{Bratos}},
  \bibinfo{author}{\bibfnamefont{J.-C.} \bibnamefont{Leicknam}},
  \bibnamefont{and} \bibinfo{author}{\bibfnamefont{S.}~\bibnamefont{Pommeret}},
  \bibinfo{journal}{Chemical Physics} \textbf{\bibinfo{volume}{359}},
  \bibinfo{pages}{53 } (\bibinfo{year}{2009}).

\bibitem[{\citenamefont{Fortes et~al.}(2005)\citenamefont{Fortes, Wood,
  Alfredsson, Vo{\v{c}}adlo, and Knight}}]{fortes05}
\bibinfo{author}{\bibfnamefont{A.~D.} \bibnamefont{Fortes}},
  \bibinfo{author}{\bibfnamefont{I.~G.} \bibnamefont{Wood}},
  \bibinfo{author}{\bibfnamefont{M.}~\bibnamefont{Alfredsson}},
  \bibinfo{author}{\bibfnamefont{L.}~\bibnamefont{Vo{\v{c}}adlo}},
  \bibnamefont{and} \bibinfo{author}{\bibfnamefont{K.~S.}
  \bibnamefont{Knight}}, \bibinfo{journal}{Journal of Applied Crystallography}
  \textbf{\bibinfo{volume}{38}}, \bibinfo{pages}{612} (\bibinfo{year}{2005}).

\bibitem[{\citenamefont{Koyama et~al.}(2004)\citenamefont{Koyama, Tanaka, Gao,
  and Zeng}}]{koyama04}
\bibinfo{author}{\bibfnamefont{Y.}~\bibnamefont{Koyama}},
  \bibinfo{author}{\bibfnamefont{H.}~\bibnamefont{Tanaka}},
  \bibinfo{author}{\bibfnamefont{G.}~\bibnamefont{Gao}}, \bibnamefont{and}
  \bibinfo{author}{\bibfnamefont{X.~C.} \bibnamefont{Zeng}},
  \bibinfo{journal}{J. Chem. Phys.} \textbf{\bibinfo{volume}{121}},
  \bibinfo{pages}{7926} (\bibinfo{year}{2004}).

\bibitem[{\citenamefont{Str\"assle et~al.}(2005)\citenamefont{Str\"assle,
  Klotz, Loveday, and Braden}}]{strassle05}
\bibinfo{author}{\bibfnamefont{T.}~\bibnamefont{Str\"assle}},
  \bibinfo{author}{\bibfnamefont{S.}~\bibnamefont{Klotz}},
  \bibinfo{author}{\bibfnamefont{J.~S.} \bibnamefont{Loveday}},
  \bibnamefont{and} \bibinfo{author}{\bibfnamefont{M.}~\bibnamefont{Braden}},
  \bibinfo{journal}{J. Phys.: Condens. Matter} \textbf{\bibinfo{volume}{17}},
  \bibinfo{pages}{S3029} (\bibinfo{year}{2005}).

\bibitem[{\citenamefont{Dunaeva et~al.}(2010)\citenamefont{Dunaeva, Antsyshkin,
  and Kuskov}}]{dunaeva10}
\bibinfo{author}{\bibfnamefont{A.}~\bibnamefont{Dunaeva}},
  \bibinfo{author}{\bibfnamefont{D.}~\bibnamefont{Antsyshkin}},
  \bibnamefont{and} \bibinfo{author}{\bibfnamefont{O.}~\bibnamefont{Kuskov}},
  \bibinfo{journal}{Solar System Research} \textbf{\bibinfo{volume}{44}},
  \bibinfo{pages}{202} (\bibinfo{year}{2010}).

\bibitem[{\citenamefont{Giauque and Stout}(1936)}]{giauque36}
\bibinfo{author}{\bibfnamefont{W.~F.} \bibnamefont{Giauque}} \bibnamefont{and}
  \bibinfo{author}{\bibfnamefont{J.~W.} \bibnamefont{Stout}},
  \bibinfo{journal}{Journal of the American Chemical Society}
  \textbf{\bibinfo{volume}{58}}, \bibinfo{pages}{1144} (\bibinfo{year}{1936}).

\bibitem[{\citenamefont{Smith et~al.}(2007)\citenamefont{Smith, Lang, Liu,
  Boerio-Goates, and Woodfield}}]{smith07}
\bibinfo{author}{\bibfnamefont{S.~J.} \bibnamefont{Smith}},
  \bibinfo{author}{\bibfnamefont{B.~E.} \bibnamefont{Lang}},
  \bibinfo{author}{\bibfnamefont{S.}~\bibnamefont{Liu}},
  \bibinfo{author}{\bibfnamefont{J.}~\bibnamefont{Boerio-Goates}},
  \bibnamefont{and} \bibinfo{author}{\bibfnamefont{B.~F.}
  \bibnamefont{Woodfield}}, \bibinfo{journal}{J. Chem. Phys.}
  \textbf{\bibinfo{volume}{39}}, \bibinfo{pages}{712 } (\bibinfo{year}{2007}).

\bibitem[{\citenamefont{Feistel and Wagner}(2006)}]{feistel06}
\bibinfo{author}{\bibfnamefont{R.}~\bibnamefont{Feistel}} \bibnamefont{and}
  \bibinfo{author}{\bibfnamefont{W.}~\bibnamefont{Wagner}},
  \bibinfo{journal}{J. Phys. Chem. Ref. Data} \textbf{\bibinfo{volume}{35}},
  \bibinfo{pages}{1021} (\bibinfo{year}{2006}).

\end{thebibliography}

\end{document}